\documentclass[10pt]{iopart}

\usepackage{iopams}  
\usepackage{graphicx}
\usepackage{natbib}

\renewcommand{\vec}[1]{\boldsymbol{#1}}

\begin{document}

\review[In-situ measurements of space plasma]{}{In-situ measurements of space plasma: recent progress and future challenges}

\author{D Verscharen$^1$}

\address{$^1$ Mullard Space Science Laboratory, University College London, Holmbury St Mary, Dorking, RH5\,6NT, UK}
\ead{d.verscharen@ucl.ac.uk}

\vspace{10pt}
\begin{indented}
\item[]21 August 2026
\end{indented}

\begin{abstract}
Space plasmas like the solar wind or the Earth's space environment offer unique opportunities to observe fundamental plasma processes and their impact in situ. With modern space instrumentation, we measure the velocity distribution function of the plasma particles as well as the electromagnetic fields at high resolution and with minimal perturbation of the observed plasma systems. Plasma measurements like this are often not possible in laboratory settings on Earth. This review article focuses on modern diagnostic methods for the in-situ detection of plasma particles in space. It presents the detection principle of top-hat electrostatic analysers and highlights recent examples of scientific discoveries based on data from the heliospheric space missions Parker Solar Probe and Solar Orbiter. These examples demonstrate the capabilities of modern space plasma instrumentation. The article then discusses future directions in space plasma physics as well as the involved challenges in terms of the required plasma diagnostics. These new developments include, for example, upcoming and proposed space missions such as the operational space-weather mission Vigil, the multi-spacecraft mission HelioSwarm, the Mars mission M-MATISSE, and the electron-astrophysics mission Debye.
\end{abstract}
%
\submitto{\PPCF}
%
%
%

\section{Introduction}

Space plasmas are the only astrophysical plasma systems that are accessible for direct in-situ measurements \citep{verscharen21,tsurutani23}. Therefore, they allow us to observe plasma processes in great detail with the help of spacecraft and scientific space instrumentation \citep{pfaff98,pfaff98b,mostafavi24b}. In this way, space plasmas serve us as astrophysical laboratories for the study of plasma physics \citep{longley22,sorriso23}. The characteristic plasma scales are often large compared to the typical dimensions of spacecraft, so that perturbations of the plasma due to our measurement devices are often negligible. Unlike plasma laboratories on Earth, however, we have very limited control over the plasma parameters in space; instead, we must generally rely on nature to provide us with the relevant plasma conditions for our scientific investigations.

It is convenient that the solar system is filled with plasma that exhibits a wide diversity of plasma conditions. These plasmas include the strongly magnetised and collisional lower solar atmosphere, partially ionised and collisional planetary ionospheres, the fully ionised and collisionless solar wind, as well as energetic plasma populations in planetary magnetospheres and radiation belts. By sending spacecraft to these different environments, we are able to explore a broad range of plasma conditions and a large variety of plasma processes in space, and even in a given space-plasma system, the conditions vary significantly in space and time. 

The solar wind at a heliocentric distance of 1 astronomical unit (au) is such a variable space plasma \citep{wilson18,verscharen19,salem23}. It expands with typical flow velocities of about 500\,km/s; however, this number varies from slow solar wind at speeds of 300 to 400\,km/s  to fast solar wind with speeds of up to about 800\,km/s. Coronal mass ejections can reach bulk velocities in excess of 2,500\,km/s. The magnetic field strength at 1\,au is typically of order a few nano-Tesla, while the plasma temperature is of order $10^5$\,K, slightly lower than the temperature of the solar corona of about $10^6$\,K from where the solar wind emerges. Due to the collisionless nature of the solar wind, electrons, protons, and other ion species often exhibit different temperatures, temperature anisotropies, and different bulk speeds \citep{marsch06}. The typical proton number density is about 3 particles per cubic centimetre, but this number also varies significantly and anti-correlates, on average, with wind speed. Combining the relevant plasma parameters leads to a mean plasma-$\beta$ of order unity at 1\,au. However, this value also varies significantly over time, giving us statistically access to plasma with $\beta$ of just a few per cent up to values of about 100. The collisional mean free path for protons at a distance of 1\,au is typically of order 1\,au, indicating that the solar wind is indeed a collisionless plasma. However, due to its expansion history starting in the collisional solar corona and considering that thermal electrons have higher collision frequencies than protons, collisional processes still impact the physics of the solar wind \citep{bale13,heidrich20,bercic21c,johnson24,mostafavi24}. Therefore, the solar wind allows us to explore the transition from (weakly) collisional to collisionless plasma conditions.

In this review article, I introduce the measurement principles used in modern space-plasma missions for the detection of plasma particles. I focus on measurements of the velocity distribution functions of particles, as space plasma diagnostics offer levels of detail and resolution in their measurements that are unachievable with laboratory plasma diagnostics on Earth. For demonstration purposes, I show examples of kinetic plasma science achieved by these measurements in the solar wind. I close by presenting some future space-plasma missions, promising mission concepts with plasma instrumentation, and the associated challenges for plasma diagnostics.

\section{Particle measurements with electrostatic analysers}

Due to the collisionless nature of many space plasmas, it is of particular interest to measure the velocity distribution function of the plasma particles rather than just the velocity moments (density, bulk velocity, pressure, etc.) of the distribution \citep{marsch06,wilson22}. Electrostatic analysers enable us to resolve the velocity distribution by counting plasma particles in bins over energy and look direction \citep{carlson82,johnstone85,fazakerley98}. These instruments typically measure particles with energies from a few eV up to 10s of keV.

\begin{figure}[h]
 \begin{center}
 \includegraphics[width=0.8\textwidth, angle=0]{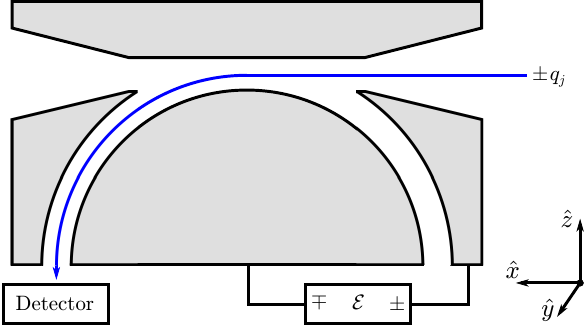}
 \end{center}
 \caption{Working principle of a top-hat electrostatic analyser. The blue curve represents the trajectory of a particle with charge $q_j$ that enters the electrostatic analyser. The grey areas represent the mechanical structure of the instrument head. A variable voltage $\mathcal E$ can be applied between the inner and outer hemispheres. Particles that fulfil Eq.~(\ref{deflection}) reach the detector at the bottom of the hemispherical analyser head. From \citet{verscharen19}.}\label{fig_esa}
\end{figure}

Figure~\ref{fig_esa} illustrates the working principle of modern top-hat electrostatic analysers. They are often built in spherical symmetry around the $z$-axis defined in this figure, leading to the geometrical association with the shape of a top hat. The grey areas in Figure~\ref{fig_esa} represent conducting metal structures that are electrically isolated from each other and from the spacecraft body.  A high-voltage power supply combined with a high-voltage modulator applies a variable voltage $\mathcal E$ between the two nested hemispheres of the detector. An incoming charged particle of species $j$ with energy $W_j$, charge $q_j$, and mass $m_j$ is deflected by the electric field associated with the voltage $\mathcal E$. The electric field is, to first order, perpendicular to the hemispheres and thus radial at each point of the particle trajectory, and its magnitude is $\propto \mathcal E/\Delta r$, where $\Delta r$ is the distance between the inner and outer hemispheres. Assuming the particle follows a circular trajectory of radius $r$ in the $x$-$z$ plane, the electric force acts as the centripetal force on the particle trajectory. The particle then only reaches the detector at the bottom of the hemispheres if it matches the condition
\begin{equation}\label{deflection}
\frac{W_j}{q_j}=\frac{1}{2}\left(\frac{r}{\Delta r}\right)\mathcal E=k\mathcal E,
\end{equation}
where $k$ is the so-called $k$-factor that is defined by the geometrical design of the detector. Many contemporary electrostatic analysers are designed towards a $k$-factor of order 10, and calibration is generally required to characterise the effective $k$-factor of any built instrument.

According to Eq.~(\ref{deflection}), for a given setting of $\mathcal E$, only particles within a range of energy-per-charge values $W_j/q_j$ arrive at the detector. The width of this energy-per-charge range is determined by the geometry of the detector, including the aperture size and the separation between both nested hemispheres. By stepping along a pre-defined sweep profile of $\mathcal E$-values while counting the incoming particles at the detector, the electrostatic analyser builds up a $W_j/q_j$-spectrum of the incoming particles. This measurement principle works under the assumption that the overall sampling time is short compared to the timescale associated with changes in the plasma distribution. 

Depending on the sign of $\mathcal E$, electrostatic analysers can detect positively or negatively charged particles; however, they are unable to distinguish different particle populations as long as they have the same $W_j/q_j$. For instance, in a plasma consisting of $\alpha$-particles and protons, the electrostatic analyser would indistinguishably register an $\alpha$-particle and a proton at the same setting of $\mathcal E$ if the $\alpha$-particle has $1/\sqrt{2}$ times the speed of the proton \citep{zhang24}. A separation by species is possible by combining the electrostatic analyser with additional devices, such as time-of-flight spectrometers \citep{gloeckler85,moore95}, or through the application of statistical methods during the analysis of the recorded count data \citep{steinberg96,nicolaou22,demarco23}.

Micro-channel plates (MCPs) and channel electron multipliers (CEMs) are commonly used detector technologies in top-hat electrostatic analysers, as they efficiently amplify the signal of the incoming particles to a level that then enables electronic amplification and counting \citep{baumgartner76,wiza79,funsten15,gershman16}. 
By making the detector sensitive to the spatial position of the incoming particles in the $x$-$y$ plane, the instrument resolves the azimuth angle $\phi$ of the incoming particles, i.e., the angle between the projection of their velocity vector onto the $x$-$y$ plane and the $x$-axis in Figure~\ref{fig_esa}. In combination with position-sensitive anodes, MCPs provide azimuth resolutions of a few degrees.

Modern top-hat electrostatic analysers are often combined with aperture deflector plates near the point where particles enter the instrument \citep{kasper16,owen20}. Like the hemispheres, the aperture deflector plates can be biased with an electric voltage and thus bend trajectories of particles with different elevation angles $\theta$ to enter the hemispherical energy-per-charge selection unit of the instrument. The elevation angle $\theta$ is defined as the angle between the particle velocity vector and the $x$-$y$ plane.  The electrostatic sampling of elevations through aperture deflectors is particularly important for instruments on three-axis-stabilised spacecraft, as we cannot rely on spacecraft spin to resolve the third dimension of the velocity distribution function. These elevation-sensitive electrostatic analysers sweep through pre-defined voltage profiles for both the hemispheres and the aperture deflector plates, and thus construct count maps $C$ of the incoming particles binned by $\mathcal E$, $\phi$, $\theta$, and time.

\begin{figure}[h]
 \begin{center}
 \includegraphics[width=0.45\textwidth, angle=0]{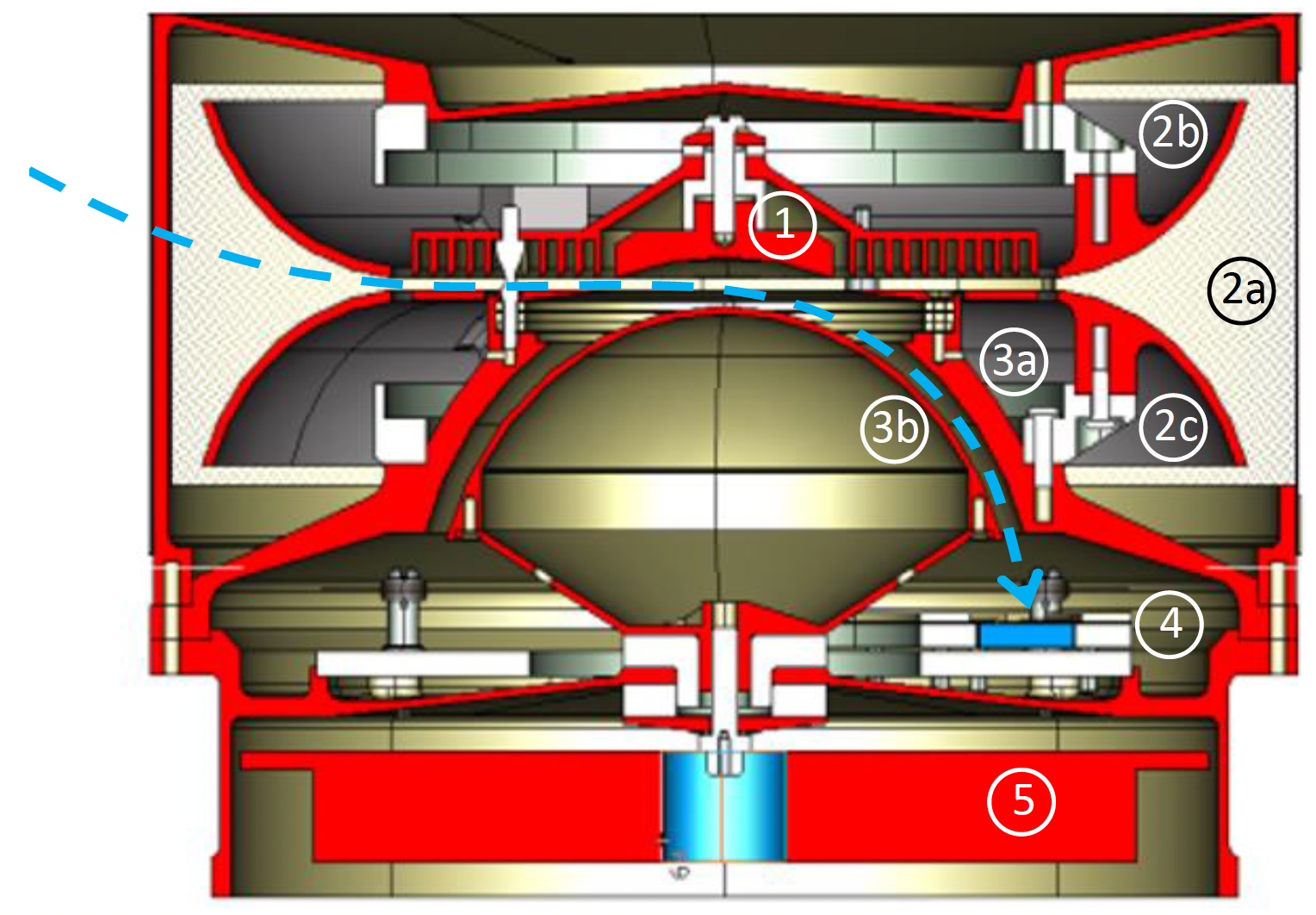}
 \includegraphics[width=0.45\textwidth, angle=0]{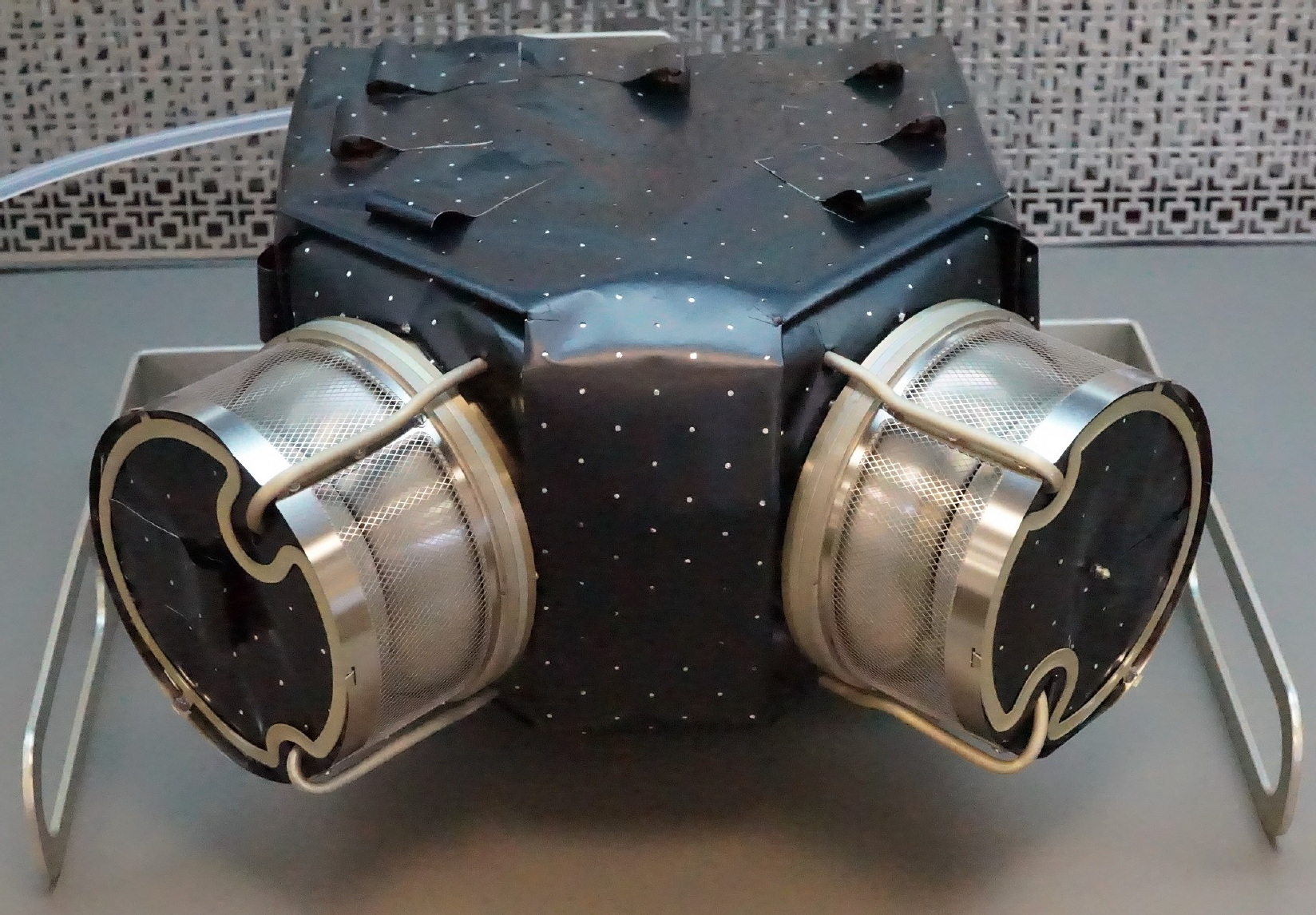}
 \end{center}
 \caption{Solar Orbiter's EAS. Left: cross-section of an EAS sensor head. The blue dashed curve indicates the trajectory of an incoming electron. The numbered labels relate to the description in the text. Right: Photograph of the EAS flight model. The two sensor heads are mounted on a common electronics box, which is covered in black multi-layer insulation. Credit: \citet{owen20}, A\&A, 642, A16, 2020, reproduced with permission \textcopyright ESO.}\label{fig_EAS}
\end{figure}

Figure~\ref{fig_EAS} shows the cross-section of one head of the Solar Wind Analyser's (SWA) Electron Analyser System (EAS) onboard the Solar Orbiter mission \citep{owen20}. This top-hat electrostatic analyser is designed to detect electrons from energies of a few eV to 5\,keV in the solar wind. An incoming electron, represented by the blue dashed line in Figure~\ref{fig_EAS}, first crosses a grounded cylindrical grid (2a), which suppresses electrostatic interference with the other instruments on the spacecraft. It then experiences electrostatic deflection through the upper (2b) and lower (2c) aperture deflector plates that select the elevation angle $\theta$ of the incoming particle. After passing through the elevation-selection stage, the particle enters the energy-per-charge selection unit, in which it experiences electrostatic deflection between the outer (3a) and inner (3b) hemispheres. If the particle matches the condition in Eq.~(\ref{deflection}), it arrives at the MCP (4), which then amplifies the signal through a cascade of secondary electrons. A position-sensitive field of anodes below the MCP picks up the signal of the electron cascade, which then enters the charge amplifiers (5) for signal processing. One unique feature of EAS is the variable-geometric-factor (VGF) system (1), which controls electrostatically the flux of incoming particles and thus the dynamic range of the instrument \citep{collinson10}.

We now derive the expected count map $C$ that a top-hat electrostatic analyser records when the detected plasma particles of species $j$ have the velocity distribution function $f_j(\vec x,\vec v,t)$.  We assume that the distribution in the spacecraft reference frame is constant in time over the timescale of the sampling of our instrument. For the sake of simplicity, we also assume that only particles of one species with known $q_j$ and $m_j$ are present, allowing us to relate their $W_j/q_j$ directly to their speed $v=\sqrt{2W_j/m_j}$.
At each pair of voltage settings on the hemispheres and the aperture deflector plates, the detector electronics count particles for a constant time $\Delta \tau$.  For each voltage setting, our instrument accepts particles in a range of speed $v$, azimuth $\phi$, and elevation $\theta$ around the central values of speed 
\begin{equation}\label{eq_UE}
U=\sqrt{\frac{2q_jk\mathcal E}{m_j}},
\end{equation}
 azimuth ($\Phi$), and elevation ($\Theta$) associated with a given bin of $C$. 
We account for the finite resolution of our instrument in energy-per-charge, azimuth, and elevation by defining non-zero acceptance widths in speed ($\Delta U$), azimuth ($\Delta \Phi$), and elevation ($\Delta \Theta$). 
With these assumptions, the number of detected particles at a defined bin in $U$, $\Phi$, and $\Theta$ is given by \citep{kessel89,cara17,nicolaou20}
\begin{eqnarray}\label{analyser_eq}
C(\mathcal E,\Phi,\Theta,t)&=&\int_{U-\Delta U/2}^{U+\Delta U/2} \int_{\Phi-\Delta \Phi/2}^{\Phi+\Delta \Phi/2} \int_{\Theta-\Delta \Theta/2}^{\Theta+\Delta \Theta/2} \int_{t-\Delta \tau/2}^{t+\Delta \tau/2} f_j(\vec x,\vec v,t^{\prime})\\  \nonumber
 & & \times \vphantom{\int\limits_A^A}A_{\mathrm{eff}}  v^3\,\mathrm dv \cos\theta\,\mathrm d\theta\,\mathrm d\phi\,\mathrm dt^{\prime},
\end{eqnarray}
where $A_{\mathrm{eff}}$ is the effective aperture of the instrument, which we assume to be a function of $\theta$ only: $A_{\mathrm{eff}}=A_0/\cos\theta$. Approximating Eq.~(\ref{analyser_eq}) through the mid-point rule and taking $f_j$ to be a function of $\vec v$ only allows us to re-write the expression for the count map $C$ as
\begin{equation}\label{Ccount}
C(\mathcal E,\Phi,\Theta,t)\approx G f_j(U,\Phi,\Theta) U^4\,\Delta \tau,
\end{equation}
where
\begin{equation}
G=A_0\frac{\Delta U}{U}\,\Delta \Phi\,\Delta \Theta
\end{equation}
is the geometric factor of the instrument. The geometric factor is often assumed to be a constant scalar; however,  it is more appropriate to evaluate and characterise $G$ as a function of $U$, $\Phi$, and $\Theta$ in calibration campaigns. Eq.~(\ref{Ccount}) has the advantage that it can be easily inverted to
\begin{equation}\label{analyser_eq2}
f_j(U,\Phi,\Theta)\approx \frac{C(\mathcal E,\Phi,\Theta,t)}{G  U^4\,\Delta \tau},
\end{equation}
where $U$ and $\mathcal E$ are linked through Eq.~(\ref{eq_UE}). Eq.~(\ref{analyser_eq2}) is the key equation to derive the distribution function $f_j$ based on the count map $C$ recorded by electrostatic analysers. At this point, we reiterate that analysers of this type register particles of $W_j/q_j$ instead of $U$, so that particles of different species may appear in the same $U$-bin even if they have indeed different velocities $v$.

Spacecraft charging can strongly affect in-situ particle measurements in space \citep{lavraud16,bergman20}. When a spacecraft is exposed to sunlight, especially to the ultraviolet spectral range, photo-electrons are emitted from the spacecraft body. A current balance is established between the photo-electron current and the plasma current to the spacecraft, such that the spacecraft assumes a floating potential with respect to the plasma \citep{whipple81,scudder00}. In the solar wind, typical spacecraft floating potentials of about +10\,V are to be expected  \citep{wilson23}.  When spacecraft enter dense plasma environments, for example in planetary magnetospheres, or experience occultations from sunlight, the spacecraft potential often assumes a negative value of the same order of magnitude. Measurements of plasma particles with an absolute energy-per-charge $|W_j/q_j|$ comparable to or below the magnitude of the spacecraft potential suffer from spacecraft charging. Since most ion measurements are taken above approximately 100\,eV-per-charge, they are less affected by spacecraft charging. However, the measurement of electrons with energies of $\lesssim 10$\,eV often suffers from charging effects. If the spacecraft is positively charged, electrons are accelerated towards the spacecraft body and the recorded energy spectrum must be corrected \citep{scime94}. If the spacecraft is negatively charged, electrons with an energy-per-charge below the magnitude of the spacecraft potential are repelled and thus not detectable \citep{song97}. Moreover, the photo-electrons emitted by the spacecraft  contaminate the measurement of plasma electrons at low energies \citep{salem01,stverak26}. Possible mitigations for the impact of spacecraft charging include the choice of material with a high work function \citep{berry81}, the minimisation of differential spacecraft charging through common grounding of most spacecraft elements, and the use of Active Spacecraft Potential Control \citep[ASPOC; ][]{riedler97,torkar16}.

For the achievement of high-precision measurements, corrections for errors due to finite counting statistics, counting noise, non-uniform acceptance widths, and other instrumental effects must be applied to the count map $C$ \citep{wilson15,nicolaou23,nicolaou24,nicolaou25}.

\section{Kinetic plasma science in the solar wind}

After the introduction of the measurement principle of electrostatic analysers, this section presents two recent examples that showcase the capabilities of these instruments for in-situ particle measurements in the solar wind. 

\subsection{High-cadence electron observations}\label{sect_laura}

Solar Orbiter is a space mission led by the European Space Agency (ESA) with the goal of understanding the links between the Sun and the heliosphere \citep{mueller20}. It was launched in February 2020, and its orbit brings the spacecraft to heliocentric distances between $\lesssim$0.3\,au and $\gtrsim$1\,au. The payload consists of ten instruments, including six remote-sensing instruments to observe the Sun and four in-situ instruments to measure the properties of the particles and fields at the location of the spacecraft. Solar Orbiter leaves the plane of the ecliptic and will achieve an inclination of over 30$^{\circ}$ during its extended mission phase, enabling an unprecedented view on the poles of the Sun \citep{garcia21}.

The EAS is part of the SWA instrument suite on board Solar Orbiter \citep{owen20}. It consists of two identical top-hat electrostatic analyser heads that are mounted on a common electronics box at an angle of 90 degrees with respect to each other (see Figure~\ref{fig_EAS}). Each head resolves 360$^\circ$ of azimuth in 32 steps and $\pm$45$^\circ$ of elevation in 16 steps. Together and in this geometrical configuration, both heads thus resolve the full sky ($4\pi$ steradian) and have a significant overlap in their individual fields of view, which is useful for cross-calibration between the heads. The energy-per-charge of the incoming particles is resolved in 65 steps with an approximately constant relative energy resolution of $\Delta W_{\mathrm e}/W_{\mathrm e}$ of about 13\%. In normal mode, the two EAS heads record their combined count maps of $2\times 65 \times 32\times 16=66\,560$ entries every second. Telemetry constraints usually do not allow us to transmit the full count maps at this high cadence though.

EAS offers a burst mode of operation \citep{owen21}. In this mode, the EAS electronics receives the current measured direction of the interplanetary magnetic field as measured by the magnetometer MAG through on-board inter-instrument communication \citep{horbury20}. The EAS electronics then selects the head in whose reference frame the magnetic-field vector has the smallest elevation angle. The aperture deflector plates of this head then steer it to the elevation angle that includes the direction of the magnetic field, and the head records the count map for all 65 energy steps and 32 azimuth steps at this elevation setting. Then the aperture deflector plates steer the head to the elevation setting that includes the direction opposite to the magnetic field, and again all 65 energy steps and 32 azimuth steps are recorded. The resulting count map of $65\times32\times2=4,160$ entries samples the full pitch-angle distribution $f_{\mathrm e}$ of the electrons, albeit recorded at different gyro-phases of the incoming particles.  Under the assumption of gyrotropy, this burst-mode measurement principle allows us to sample electron pitch-angle distributions at a high rate of 8 distributions per second.

\begin{figure}[h]
 \begin{center}
 \includegraphics[width=\textwidth, angle=0]{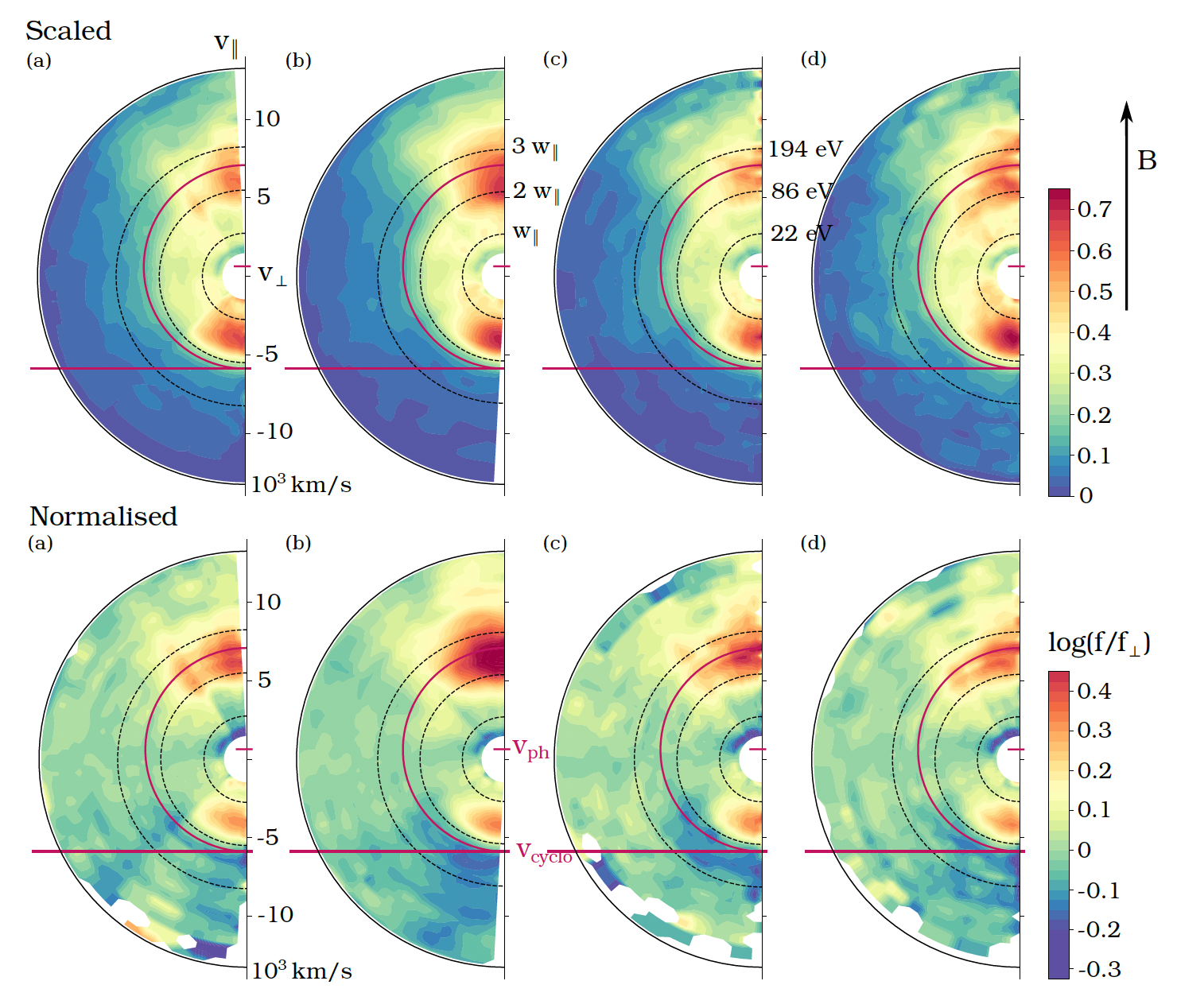}
 \end{center}
 \caption{Solar Orbiter EAS burst-mode observations of the electron distribution function. The top row shows scaled and the bottom row shows normalised distributions. The vertical axis represents the velocity component $v_{\parallel}$ parallel to the local magnetic field $\vec B$, and the horizontal axis represents the velocity component $v_{\perp}$ perpendicular to $\vec B$.  The four columns represent different time instances. The distributions show various anisotropic non-equilibrium features. Credit: \citet{bercic21}, A\&A, 656, A31, 2021, reproduced with permission \textcopyright ESO.From \citet{bercic21}.}\label{fig_bercic1}
\end{figure}

Figure~\ref{fig_bercic1} shows example measurements of burst-mode distributions recorded by EAS in the solar wind at a heliocentric distance of 0.52\,au \citep{bercic21}. The top row shows the scaled electron distribution, for which the values of $f_{\mathrm e}$ in each ring of constant energy are scaled to values between 0 and 1. The bottom row shows the normalised electron distribution, for which the values of $f_{\mathrm e}$ in each ring of constant energy are normalised to the value of $f_{\mathrm e}$ at $v_{\parallel}=0$, where $v_{\parallel}$ is the velocity component parallel to the local magnetic field $\vec B$, which is measured by MAG. The scaling and normalising of the electron distribution emphasise departures of isotropy in different ways. The four columns represent different instances in time during the given solar-wind interval. The coordinate system is aligned with the direction of the local magnetic field $\vec B$, so that the vertical axes represent $v_{\parallel}$ and the horizontal axes represent the velocity component $v_{\perp}$ perpendicular to $\vec B$.

The measured electron distributions exhibit clear non-equilibrium features. For instance, at speeds above $5\times 10^3$\,km/s -- corresponding to approximately two thermal speeds $w_{\parallel}$ and energies above 80\,eV -- an enhancement in $f_{\mathrm e}$ at small pitch-angles (i.e., aligned with $\vec B$) is visible. This suprathermal component of the electron population consists of particles that escape the heliosphere and form the so-called ``strahl'' population of the solar-wind electrons \citep{feldman75,pilipp87,lin98}. This field-aligned and anti-Sunward beam of electrons undergoes adiabatic focusing by the mirror force in the decreasing interplanetary magnetic field.

Moreover, we observe a deficit in $f_{\mathrm e}$ at $v_{\parallel}<0$ near the purple horizontal line in Figure~\ref{fig_bercic1}. This deficit on the Sunward side of the distribution is a result of particle escape in the heliosphere \citep{lemaire71,maksimovic01,bercic21b,halekas21}. The fast strahl electrons that escape the heliosphere overcome the interplanetary electrostatic potential and thus do not return towards the Sun, unlike electrons at lower energies. Therefore, this lack of reflected and returning high-energy electrons appears as a Sunward deficit in the distribution. 

\citet{bercic21} demonstrate that the Sunward deficit can serve as a source of free energy for the excitation of a linear micro-instability that drives fast-magnetosonic/whistler (FM/W) waves unstable. Particles with $v_{\parallel}\approx v_{\mathrm{cyclo}}$, which is indicated by the purple horizontal lines in Figure~\ref{fig_bercic1}, can undergo cyclotron resonance with parallel-propagating FM/W waves of phase speed $v_{\mathrm ph}$. In this wave--particle interaction, the resonant particles scatter into the deficit, lose kinetic energy, and thus lead to the growth of the resonant FM/W waves.  This interaction modifies the shape of the electron distribution and thus reduces the electron heat flux \citep{coburn24}.

\begin{figure}[h]
 \begin{center}
 \includegraphics[width=\textwidth, angle=0]{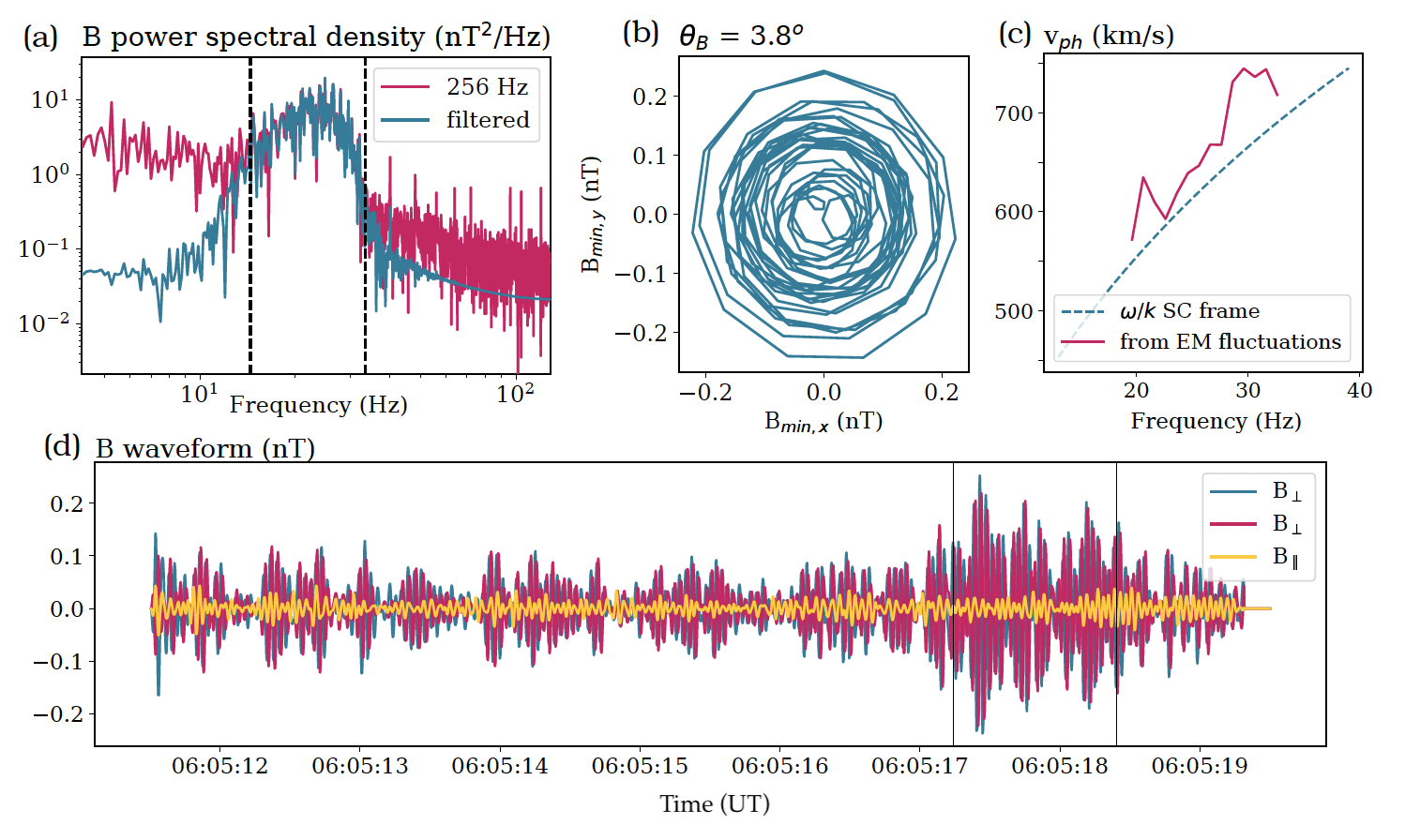}
 \end{center}
 \caption{Solar Orbiter RPW observations of the vector magnetic field during the time interval over which the electron distributions in Figure~\ref{fig_bercic1} are recorded.  (a) Power spectral density. (b) Hodogram of the magnetic-field vector. (c) Mapping of the recorded frequency in the spacecraft frame to the phase speed $v_{\mathrm{ph}}$ in the plasma frame. (d) Timeseries of the magnetic-field waveform. The measurements reveal clear signatures of FM/W waves. Credit: \citet{bercic21}, A\&A, 656, A31, 2021, reproduced with permission \textcopyright ESO.From \citet{bercic21}.}\label{fig_bercic2}
\end{figure}
Solar Orbiter's Radio and Plasma Waves (RPW) instrument records AC electric and magnetic fields \citep{maksimovic20}. At the same time when EAS detects the sporadic occurrences of the Sunward electron deficit, RPW records evidence for the presence of parallel-propagating FM/W waves. Figure~\ref{fig_bercic2} presents a summary of these wave measurements. Panel (a) shows an enhancement in the magnetic-field power spectral density at the frequencies that correspond to the resonant waves with the matching phase speed $v_{\mathrm ph}$. Panel (b) depicts the hodogram of the magnetic-field vector and highlights the right-hand circular polarisation of the waves. Panel (c) compares $v_{\mathrm ph}$ with the measured frequency of the magnetic-field signal in the spacecraft reference frame, which is needed for the mapping of the resonance condition to $v_{\mathrm{cyclo}}$ in Figure~\ref{fig_bercic1}. Panel (d) shows a timeseries of the magnetic-field waveform signal, illustrating the sub-intervals of enhanced FM/W-wave activity. The magnetic-field observations are consistent with the expectation that the Sunward deficit drives an instability of parallel-propagating FM/W waves. Later, this instability process has also been confirmed in particle-in-cell simulations \citep{micera25}.

These observations exemplify the capabilities of  particle (and fields) measurements with modern space  missions such as Solar Orbiter and highlight the ways in which we use space plasmas as natural laboratories to study the physics of fundamental plasma processes.

High-resolution electron measurements are also of great importance in other space environments. The Fast Auroral SnapshoT (FAST) mission, for example, has targeted such measurements in the magnetospheric auroral acceleration region \citep{carlson98}. Its measurements of wave--particle interactions in this low-$\beta$ geospace environment reveal the importance of kinetic instabilities driven by ion and electron beams \citep{mcfadden99}. Like in the case of the solar-wind electron distribution described above, these in-situ observations allow us to probe the interplay between large-scale gradients in the plasma system and local small-scale wave--particle interactions that jointly shape the  velocity distribution \citep{mutel07}. The combination of in-situ measurements of the three-dimensional velocity distribution with in-situ fields measurements provides insights into these self-consistent plasma processes to a level of detail that is not accessible in laboratory experiments on Earth \citep{delory98,bryant99}.

\subsection{Strong proton instabilities in the near-Sun solar wind}

Parker Solar Probe is a NASA-led space mission with the goal of exploring the near-Sun space environment \citep{fox16}. It was launched in August 2018, and its orbit brings the spacecraft to heliocentric distances between $\lesssim$0.05\,au and $\lesssim$1\,au. Its payload consists of four instruments, including three in-situ instruments and one coronal white-light imager.

Parker Solar Probe carries the  Solar Wind Electrons Alphas and Protons (SWEAP) instrument suite \citep{kasper16}, which includes the Faraday cup Solar Probe Cup \citep[SPC; ][]{case20} and two types of Solar Probe Analyzers \citep[SPAN; ][]{whittlesey20,livi22}. The SPANi unit is a top-hat electrostatic analyser for the measurement of ions. It is attached to the side of the spacecraft and thus protected from the Sun's radiation by the spacecraft's heat shield. Although the heat shield often reduces the field of view of SPANi, aberration due to the spacecraft motion and the thermal width of the ion distributions allow us to sample large fractions of the proton and $\alpha$-particle distributions most of the time, especially at small heliocentric distances. SPANi is equipped with a time-of-flight unit that separates proton and $\alpha$-particle counts in the data. In nominal mode, the instrument resolves energy-per-charge from 2\,eV/e to 30\,keV/e in 32 steps, 247.5$^{\circ}$ of azimuth in 8 steps, and 120$^{\circ}$ of elevation in 8 steps, although finer sweep steps can be commanded. The nominal measurement duration for a full sweep is 0.218\,s.

Parker Solar Probe's SWEAP instrument suite provides the first measurements of proton velocity distribution functions in the solar wind at heliocentric distances below 0.29\,au and thus close to the source regions and drivers of the solar wind. Its unprecedented orbit makes it the first mission with plasma instrumentation to break the heliocentric-distance record set by the Helios spacecraft in the 1970s. The electrostatic analysers on Helios sampled proton distributions with a sweep duration of 10\,s.  Proton measurements in the solar wind are challenging due to the low fluxes of particles compared to other space-plasma environments, the required high time resolution to account for the fast advection of plasma across the spacecraft, the supersonic (``cold-beam'') nature of the plasma protons, and the need to resolve non-thermal features in the tails of the distribution where count rates are generally low. The measurement performance requirements of the SWEAP suite are tailored to account for these difficulties in the environment explored by Parker Solar Probe.

\begin{figure}[h]
 \begin{center}
 \includegraphics[width=0.7\textwidth, angle=0]{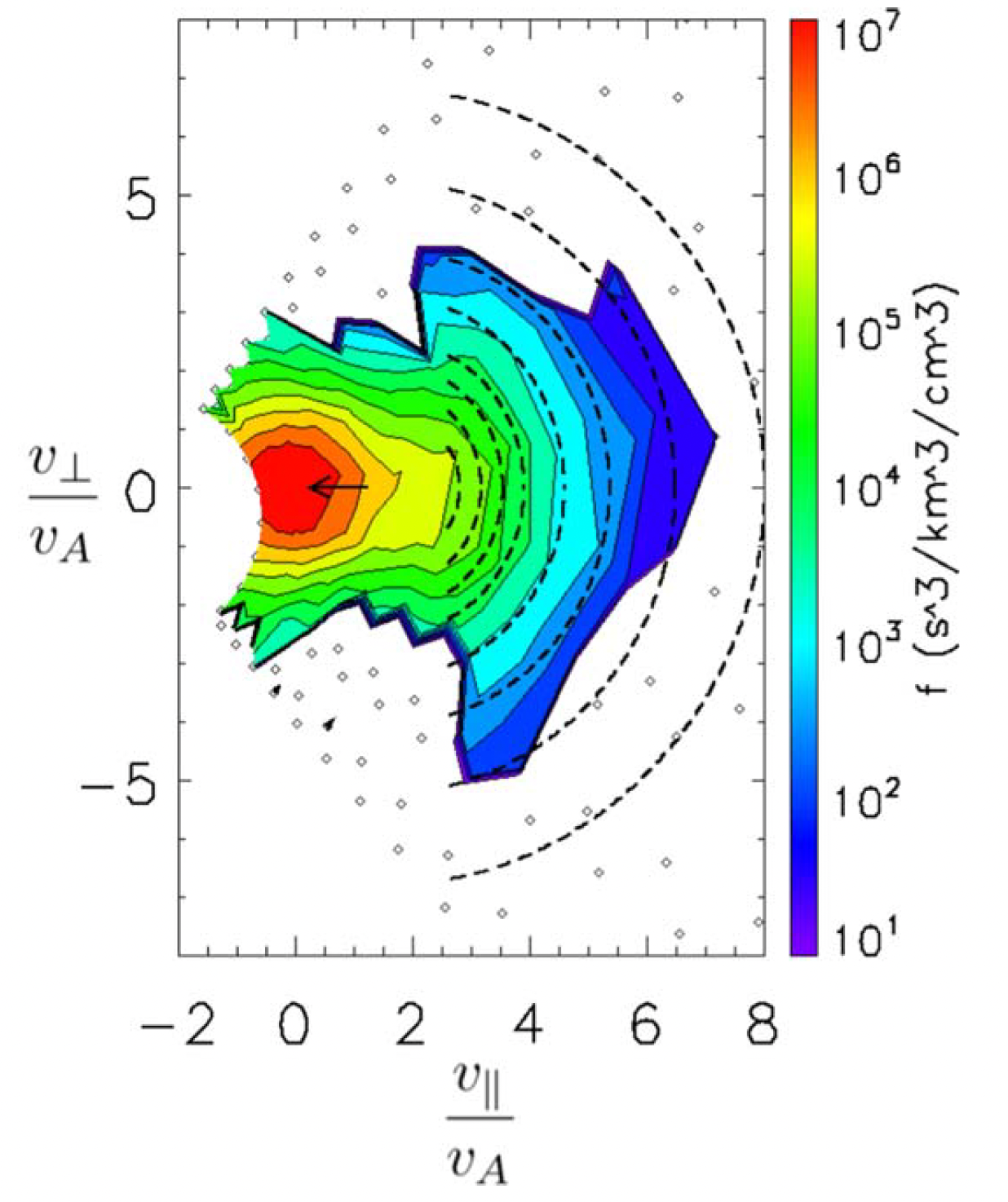}
 \end{center}
 \caption{Parker Solar Probe SPANi measurements of the proton velocity distribution function. The colours show the value of $f_{\mathrm p}$ in the instrument's $\theta$-plane in a coordinate system that is aligned with the magnetic field $\vec B$. The velocity components are normalised in units of the local Alfv\'en speed $v_{\mathrm A}$. The distribution exhibits a strong hammerhead signature in its field-aligned beam component. Reproduced from \citet{verniero22}. \textcopyright The Author(s). Published by IOP Publishing Ltd. CC BY 4.0.}\label{fig_verniero1}
\end{figure}
Figure~\ref{fig_verniero1} shows measurements of the proton velocity distribution function $f_{\mathrm p}$ recorded by SPANi at a heliocentric distance of 0.13\,au \citep{verniero22}. The colour-coding shows a cut through the three-dimensional $f_{\mathrm p}$ in the $\theta$-plane of the instrument, rotated into a coordinate system aligned with the local magnetic field $\vec B$ as measured by Parker Solar Probe's fluxgate magnetometer \citep{bale16}. The velocity components $v_{\parallel}$ and $v_{\perp}$ are normalised to the local Alfv\'en speed $v_{\mathrm A}$\footnote{The velocity component $v_{\perp}$ in Figure~\ref{fig_verniero1} assumes positive and negative values, as it represents one of the two Cartesian components of the velocity vector in the plane normal to $\vec B$. The study shown in Figure~\ref{fig_bercic1} follows a different convention in which the symbol $v_{\perp}$ represents the radial component of the cylindrical velocity coordinate system and is thus always positive.}. The grey dots indicate the centres of the measurement points of the sensor. The maximum of $f_{\mathrm p}$ is located near the bulk velocity of the plasma, indicated by the tip of the black arrow. At larger energies, $f_{\mathrm p}$ deviates from a Maxwellian equilibrium. It exhibits an extended field-aligned beam component at large $v_{\parallel}$ which shows symmetry in $v_{\perp}$ due to its gyroptropy. While proton beam populations are not uncommon in the solar wind \citep{marsch82,alterman18,durovcova21,bruno24}, the shape of this beam with its extended wings towards large $|v_{\perp}|$ is not typically seen at heliocentric distances of around 1\,au. The observed feature appears as a common structure in the solar wind close to the Sun.

Due to its shape, the observing team refer to this type of distribution as ``hammerhead distribution'' \citep{verniero20}. However, the observed shell-like structure is a known feature in plasma physics \citep{rowlands66,dusenbery81,marsch82a,isenberg96,singh98}. Kinetic wave--particle interactions through the anomalous Doppler resonance scatter resonant particles in pitch-angle and, to some degree, in energy \citep{cuperman69,parail78,ishihara84,galinsky00}. For example, proton beams can serve as sources of free energy for the excitation of a family of beam-driven  micro-instabilities that act through the anomalous Doppler resonance \citep{montgomery76,marsch87,matteini13}. Low-frequency FM/W waves can resonate with beam protons, leading to quasilinear diffusion of the resonant particles in velocity space \citep{gary84,gary85}. In this process, the particles lose kinetic energy while their $v_{\parallel}$ decreases and their $|v_{\perp}|$ increases \citep{daughton99}. This scattering process diffuses $f_{\mathrm p}$ in the range of resonant beam protons along contours in velocity space that are locally centred around the phase speed of the resonant waves \citep{hellinger11,yao20}. These contours are overplotted in Figure~\ref{fig_verniero1} as dashed circles \citep[see also][]{verscharen13}. The isocontours of $f_{\mathrm p}$ near the hammerhead feature follow these velocity-space contours to first order. This shape of the isocontours of $f_{\mathrm p}$ is thus consistent with a scenario in which anomalous Doppler-resonant wave--particle interactions diffuse proton-beam particles towards larger $|v_{\perp}|$ and create the shell-like feature in $f_{\mathrm p}$.

Distributions with similar shell-like configurations can also be excited by a family of fan instabilities, which are found, for example, in magnetospheric plasmas \citep{kozyra94,bingham10}, low-$\beta$ tokamak plasmas \citep{parail80,santini84,fuchs88,lu10}, and remote astrophysical objects \citep{kazbegi91,gogoberidze05,melrose21}. Fan instabilities act through strong pitch-angle scattering of cyclotron-resonant particles \citep{vaivads95}, like in the case of the proposed scenario for the formation of the distribution shown in the Parker Solar Probe observations. Energetic ions can also drive highly oblique modes unstable through cyclotron-resonant interactions, which then lead to strong particle scattering \citep{dendy93}.

\begin{figure}[h]
 \begin{center}
 \includegraphics[width=\textwidth, angle=0]{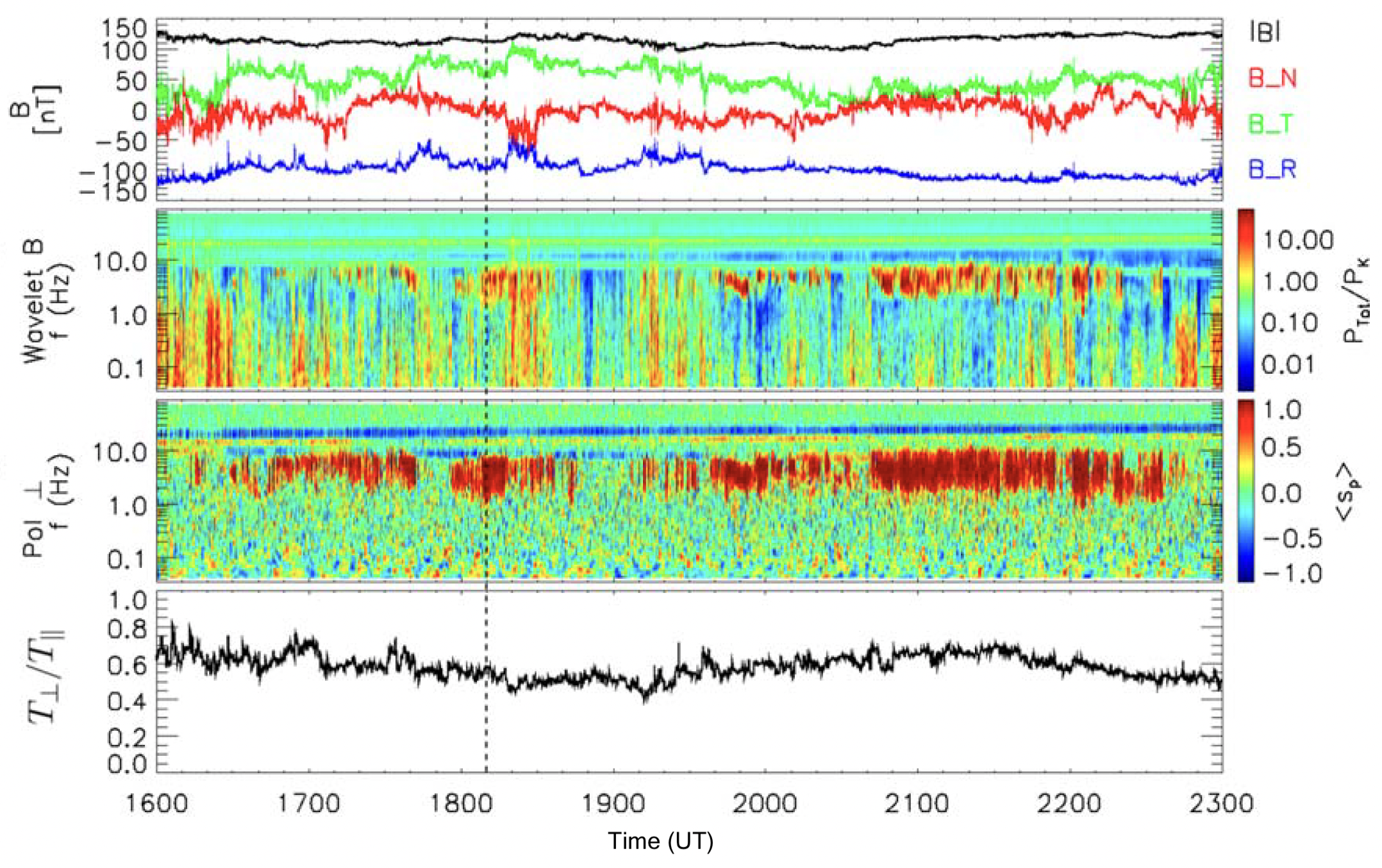}
 \end{center}
 \caption{Parker Solar Probe FIELDS measurements of the magnetic field during the time interval when SPANi records the hammerhead distribution shown in Figure~\ref{fig_verniero1}. The first panel shows the three components of the magnetic field $\vec B$ and its magnitude. The second panel shows the power-spectral density of the magnetic-field fluctuations as a function of time and frequency. The third panel shows the  perpendicular polarisation of the magnetic-field fluctuations. The fourth panel shows the temperature ratio $T_{\perp}/T_{\parallel}$ of the overall proton distribution. The FIELDS measurements reveal the presence of FM/W waves in the frequency range associated with cyclotron-resonant wave--particle interactions between protons in the hammerhead structure and these waves. Reproduced from \citet{verniero22}. \textcopyright The Author(s). Published by IOP Publishing Ltd. CC BY 4.0.}\label{fig_verniero2}
\end{figure}
Parker Solar Probe's magnetometer, which is part of the spacecraft's FIELDS instrument suite \citep{bale16}, records indeed FM/W waves at the times when significant hammerhead features are observed. Figure~\ref{fig_verniero2} shows measurements of the magnetic field from FIELDS. The first panel from the top displays the three components of $\vec B$ as well as its magnitude. The second panel shows the normalised power spectral density of magnetic-field fluctuations based on a wavelet analysis of the recorded signal. The third panel presents the frequency-dependent polarisation of the magnetic-field fluctuations. Red patches indicate a prevalence of right-handed polarisation, which is consistent with the occurrence of FM/W waves. The bottom panel shows the ratio of the proton temperatures $T_{\perp}$ perpendicular to $\vec B$ and $T_{\parallel}$ parallel to $\vec B$ measured by SPANi. Due to the strong hammerhead feature, the overall distribution $f_{\mathrm p}$ has generally an anisotropy with $T_{\perp}<T_{\parallel}$.  During times of strong hammerhead signatures in $f_{\mathrm p}$, these field measurements  consistently show the presence of coherent FM/W waves, corroborating the scenario  in which an anomalous Doppler-resonant instability fans the distribution by scattering into the shape that the observing team call the hammerhead feature.

Like in the case of the electron observations presented in Sect.~\ref{sect_laura}, these ion observations highlight the opportunities for kinetic studies in the solar wind with modern space instrumentation.

\section{Future space missions and mission concepts}

Many future space missions with in-situ plasma instrumentation are currently being developed. While some of these missions have been adopted and are preparing for launch, others are at the stage of mission concepts and proposals for future implementation. This section presents a few example cases for future missions and concepts.

\subsection{Vigil}

Vigil is an operational space-weather mission currently under development by ESA and scheduled for launch in 2031. Space weather describes the impacts of the Sun and the interplanetary plasma environment on human technology and  society \citep{schwenn06,temmer21}. In our increasingly technology-dependent society, space-weather events can have a detrimental effect on ground-based and space-based infrastructure \citep{oughton17,eastwood18,buzulukova22,miteva23}. Therefore, national governments and international organisations  increasingly recognise the need for improved space-weather forecasting capabilities, and Vigil is a major milestone towards the goal of providing these services to space-weather stakeholders in Europe and around the world  \citep{schrijver15}.

The Vigil spacecraft will be positioned at the fifth Sun--Earth Lagrange point (L5), which is a quasi-stable orbital position in the gravitational three-body system of the Sun, the Earth, and the spacecraft. L5 is located at a distance of 1\,au from the Sun and lies 60$^{\circ}$ behind the Earth in its orbit around the Sun. Due to the Sun's rotation, the L5 vantage point offers unprecedented views on regions on the surface of the Sun about 4.5 days before they become visible from Earth \citep{vourlidas15,hapgood17,rodriguez20,majirsky25}. In-situ measurements at L5 sample plasma conditions that, assuming approximately steady conditions in the solar wind's source regions, Earth will encounter likewise about 4.5 days later. Moreover, L5 offers a side view on the Sun-Earth line, which allows the tracking of plasma structures on their quasi-radial trajectory through space towards Earth \citep{akioka05,schmidt96}.

\begin{figure}[h]
 \begin{center}
 \includegraphics[width=0.6\textwidth, angle=0]{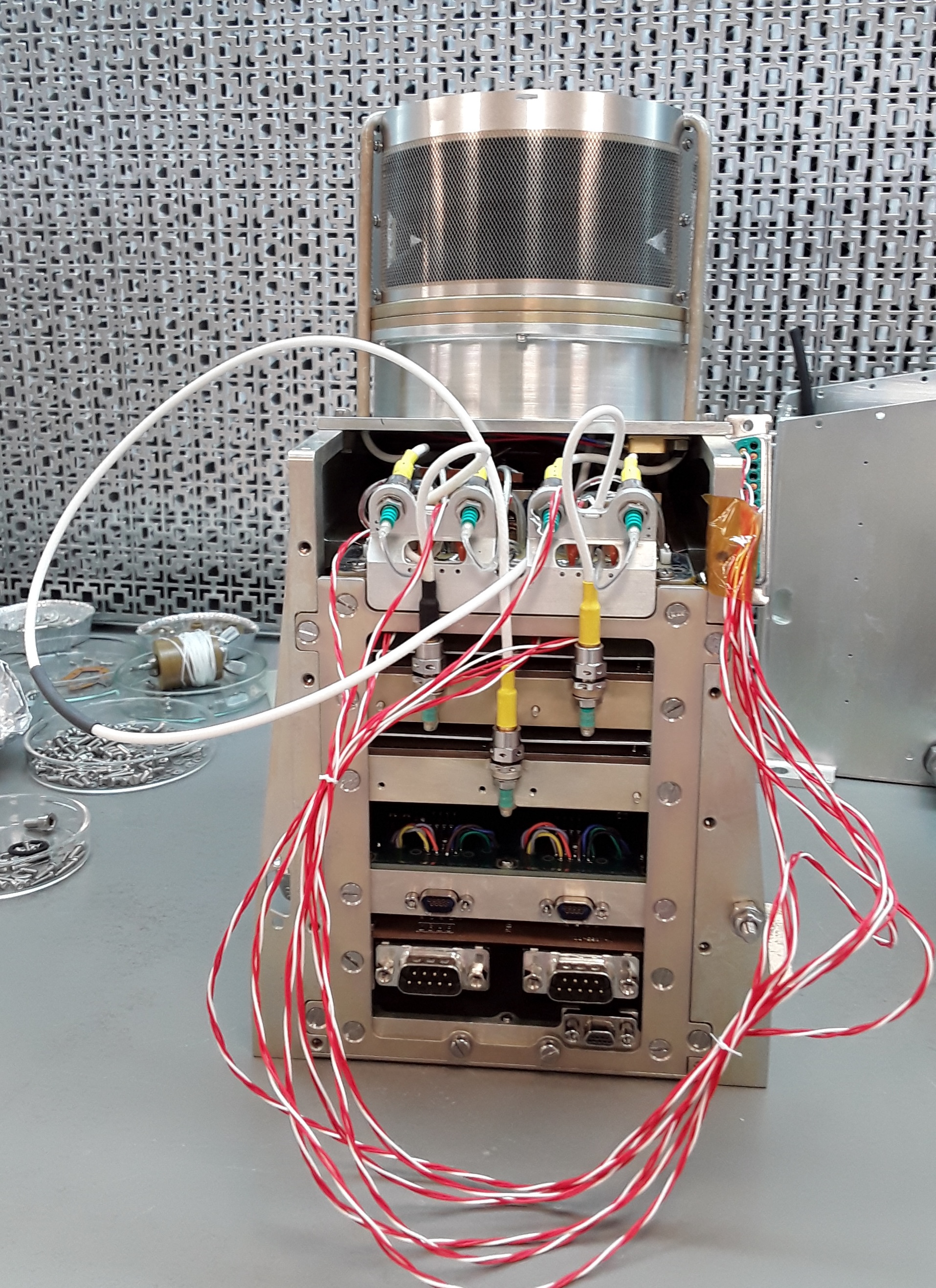}
 \end{center}
 \caption{Photograph of the Vigil PLA Demonstrator Model. The top-hat electrostatic analyser's sensor head is attached on an electronics box that contains five electronic boards. The head is surrounded by a grounded grid to avoid high-voltage interference with other instruments and the spacecraft.}\label{fig_vigil_pla}
\end{figure}

The Vigil spacecraft will carry a combined payload of two in-situ and three remote-sensing instruments:
\begin{itemize}
\item The \textbf{Plasma Analyser (PLA)} is a top-hat electrostatic analyser that will measure the protons of the solar wind at the position of the spacecraft. It builds on heritage from Solar Orbiter's EAS, but has undergone design changes to enable the measurement of protons instead of electrons \citep{nicolaou20}. PLA will measure the velocity distribution function of protons at a cadence of (at least) one distribution every minute. The ground-processing pipeline will then determine the number density, the bulk velocity, and the temperature of the plasma through the evaluation of the  velocity moments of the distribution function. Figure~\ref{fig_vigil_pla} shows the Demonstrator Model of the current PLA sensor design, including the analyser head and the electronics box. The instrument will be supported by a separate Data Processing Unit.
\item The \textbf{Magnetometer (MAG)} is a fluxgate magnetometer that will measure the direction and magnitude of the interplanetary magnetic field at L5 \citep{eastwood24}. Based on heritage from ESA's JUICE mission, MAG consists of an inboard sensor and an outboard sensor to facilitate the correction of spacecraft artifacts in the detected signals. MAG will measure field strengths between 0.1\,nT and 200\,nT on each axis at an accuracy of $\pm$1\,nT and with a cadence of (at least) 1 vector per second.
\item The \textbf{Photospheric Magnetic Field Imager (PMI)} is a vector magnetograph with heritage from Solar Orbiter's PHI instrument \citep{staub20}. As a remote-sensing instrument, it will observe the Sun's photosphere in six different wavelengths and in four polarisation states. In addition to continuum imagery of the solar surface, polarisation measurements of the Fe I line at 617.3\,nm will allow the PMI instrument to measure the photospheric line-of-sight plasma velocity through the Doppler effect and the photospheric magnetic field through the Zeeman effect. 
\item The \textbf{Heliospheric Imager (HI)} is a wide-angle white-light imager that will observe the Sun--Earth line remotely. It will provide side views at an angular resolution of 5\,arcmin that will enable the tracking of plasma structures such as coronal mass ejections from the Sun to Earth. In combination with other assets, stereoscopic observations will facilitate a reliable determination of the trajectory of these plasma structures with relevance for space weather.
\item The \textbf{Compact Coronagraph (CCOR)} is a white-light imager that will detect sunlight that is Thomson-scattered off plasma electrons in the corona and in near-Sun space. The CCOR instrument on Vigil is the third iteration of a family of compact coronagraphs that operate on other space-weather missions \citep{thernisien25}. An occulting disc inside the instrument will cover the bright solar surface and thus create an artificial solar eclipse in the instrument. This technology will allows us to observe the onset of coronal mass ejections and streamer structures in the solar corona.
\item The \textbf{Joint EUV Coronal Diagnostic Investigation (JEDI)} is a high-cadence extreme-ultraviolet (EUV) telescope that will observe the hot plasma in the solar corona out to distances of about six solar radii above the solar photosphere. It builds on heritage from Solar Orbiter's EUI instrument and PROBA-2's SWAP instrument. High-resolution images of the Sun in EUV  reveal the structure of the corona as the source region of the solar wind and other space-weather events. Flares, the release of coronal mass ejections, and coronal holes as sources of fast solar-wind streams will be visible in the EUV imagery provided by the JEDI instrument.
\end{itemize}

The main challenges for the Vigil mission result largely from its nature as an operational space-weather mission. The anticipated value of Vigil's observations for the forecasting of space weather strongly correlates with the instrumentation's high reliability, the mission's almost continuous operation, and the low latency of the data. Moreover, the Vigil mission must itself be more resilient to space-weather events than most other scientific space missions in order to provide valuable data also under extreme space-weather conditions \citep{parent24}.

Vigil complements other operational space-weather missions such as the Space Weather Follow On (SWFO) series or planned missions to the fourth Lagrange point L4 \citep{bemporad21,cho23,lee24,moon24}.

\subsection{HelioSwarm}

The HelioSwarm mission is a multi-spacecraft NASA science mission to study plasma turbulence in space plasmas \citep{klein23}. Turbulence plays a fundamental role in the transport of mass, momentum, and energy in plasma systems, and its dissipation is an important heating mechanism \citep{bruno13,chen16,schekochihin22}. Turbulence is characterised by fluctuations in the bulk velocity over a broad range of scales and a nonlinear transfer (i.e., a cascade) of fluctuation energy across scales. If the frozen-in condition is fulfilled, the magnetic field follows a fluctuation spectrum that is similar to the spectrum of the bulk-velocity fluctuations. When the cascading fluctuations reach small scales, the energy associated with the turbulent fluctuations converts to particle heat through irreversible dissipation  \citep{marino08,schekochihin09,stawarz09,coburn12}. 

The three-dimensional spatial structure and the dynamics of turbulence in weakly collisional plasmas are major scientific unknowns in our understanding of plasma physics \citep{osman07,sahraoui10,chen12,chandran15}. HelioSwarm is designed to reveal the spatial transport of energy, the transfer of energy between particles and fields, and the thermodynamic impact of intermittent structures on the plasma particles. Moreover, HelioSwarm aims to ascertain the mutual impact of turbulence near plasma boundaries, such as the Earth's bow shock and magnetosphere, and large-scale structures.

HelioSwarm is a constellation with nine spacecraft -- one hub and eight nodes -- travelling in a lunar-resonant elliptical orbit that accesses the solar wind, the foreshock, and the magnetosphere in approximately equal portions. The constellation's inter-spacecraft separations are not rigid but slowly evolve to achieve the geometries necessary for the application of the relevant analysis methods. These methods combined with HelioSwarm's multi-point space measurements provide us with simultaneous samples of spatial baselines covering the short-wavelength end of the inertial range of the turbulent fluctuations and ion scales in different plasma environments. The mission is under development for a launch in 2030.

The HelioSwarm spacecraft will carry the following instruments
\begin{itemize}
\item The \textbf{Ion Electrostatic Analyzer (iESA)} will fly on the hub spacecraft only. It builds on heritage from Solar Orbiter and MAVEN. The iESA will detect the ion velocity distribution and thus observe signatures of dissipation processes in the kinetic properties of the plasma. Its measurements will be crucial for the understanding of turbulent heating and the exchange of energy between electromagnetic fields and particles.
\item Each spacecraft will carry a \textbf{Fluxgate Magnetometer (FGM)}. The design of these instruments is a continuation of the successful line of magnetometers operating on Solar Orbiter and JUICE. The FGM will provide multi-point measurements of the vector magnetic field. It will determine the averaged background magnetic field as well as fluctuations in the magnetic field across the inertial range of the turbulence. These measurements will allow us to extract correlations of the turbulent fluctuations and to gain crucial insights into the three-dimensional structure of magnetised plasma turbulence \citep{pecora24}.
\item Each spacecraft will carry a \textbf{Search Coil Magnetometer (SCM)}. These instruments build on heritage from JUICE.  They will detect fluctuations in the vector magnetic field at high frequencies in the spacecraft frame. Therefore, the SCMs on HelioSwarm will extend the range of spatial and temporal scales in the magnetic-field measurements accessible by the FGM.
\item Each spacecraft will carry a \textbf{Faraday Cup (FC)}. These instruments build on heritage from Parker Solar Probe, Wind, and DSCOVR. Unlike the iESA, the FCs will not resolve structures in the velocity distribution of the particles, but they record the density, bulk velocity, and temperature of the plasma protons reliably and at high time resolution. Through correlation of the multi-point measurement of these plasma velocity moments, the FCs will characterise the nonlinear fluid-like interactions that are responsible for the nonlinear cascade of the plasma turbulence.
\item As a student-collaboration option, the hub spacecraft will carry an \textbf{Electron Electrostatic Analyzer (eESA)}. This instrument inherits design elements from Parker Solar Probe. It will detect signatures of kinetic processes in the electron velocity distribution function. The smallest characteristic scales of the plasma are associated with electron-kinetic processes. Therefore, electrons define the final bottleneck for the transfer of turbulent energy, making them an interesting target for extending the baseline HelioSwarm science.
\end{itemize}

The main challenges for the HelioSwarm mission largely result from its multi-spacecraft character. While the orbital design of a free-floating formation of spacecraft is complex in itself, the assembly, integration, testing, and cross-calibration of such a large number of instruments come with significant logistical and technical difficulties. 

HelioSwarm is complementary to other multi-spacecraft missions with space-plasma payload. The Cluster mission, which consisted of four spacecraft, was the first major magnetospheric multi-spacecraft mission \citep{escoubet97,goldstein15,escoubet21}. It was followed by THEMIS \citep{angelopoulos08} and MMS \citep{burch16}. In addition, the Plasma Observatory, which has won the competition for ESA's M7 call for a medium-size science mission in 2026, provides a complementary payload and orbit for the multi-scale observation of plasma energisation and energy transport \citep{retino22}.

\subsection{M-MATISSE}

The Mars Magnetosphere ATmosphere Ionosphere and Space-weather SciencE (M-MATISSE) mission is a candidate for a science mission to study the complex and dynamic coupling between Mars' magnetosphere, ionosphere, and thermosphere \citep{sanchez22}. In 2026, it has concluded its Phase-A studies in competition for ESA's M7 call for a medium-size science mission. Unlike Earth, Mars only has a minimal intrinsic magnetic field in the form of crustal magnetisation \citep{connerney01,mitchell01}. Therefore, the interactions between the solar wind and Mars differ significantly from those between the solar wind and Earth \citep{mazelle04,nagy04,halekas17,weber19}. For instance, plasma interactions around Mars, which involve the planet's ionosphere, form an induced magnetospheric boundary that deflects solar-wind plasma around the planet \citep{acuna98,bertucci11}.  Regions of strong crustal magnetisation create a hybrid magnetosphere locally that is highly variable as these regions rotate with the planet, leading to complex interactions with the solar wind. 
The dissipation of incoming energy from the solar wind has a major impact on the evolution of Mars' atmosphere and climate through atmospheric loss and through the deposition of energy via auroral processes \citep{chassefiere04,lundin04,barabash07,schneider15,schneider18}. The complex plasma interactions at Mars are largely unknown and require simultaneous observations of the upstream conditions and the magnetospheric/ionospheric response to the external drivers. 

M-MATISSE will consist of two spacecraft (called Henri and Marguerite). While Henri's orbit is designed for the spacecraft to spend most of the time within the Martian plasma system, Marguerite is intended to be located in the solar wind or the far tail of the Martian magnetosphere. Their simultaneous measurements will characterise the global dynamics of the coupling between the solar wind and the Martian magnetosphere, ionosphere, and thermosphere by unravelling the system's temporal and spatial variabilities. M-MATISSE will make a major contribution to our understanding of the Martian space environment and thus prepare for future human and robotic exploration of the planet \citep{hapgood19}.

Both M-MATISSE spacecraft will carry identical scientific instrumentation:
\begin{itemize}
\item The \textbf{Mars - Ensemble of Particle Instruments (M-EPI)} is a payload suite that will provide combined measurements of multiple particle species. Its three sensors are served by a common \textbf{Data Processing Unit (M-EPI DPU)}, which follows a design similar to the one operating on Solar Orbiter. The \textbf{Mars - Electron Analyser System (M-EAS)} is an electrostatic analyser that builds strongly on heritage from Solar Orbiter's EAS. It will measure electron distribution functions in the energy range from a few eV to 5\,keV. While telemetry constraints will not allow us to transmit full velocity distributions at a high cadence, M-EAS will extract pitch-angle distributions and partial distributions at a cadence of one per 4\,s. These data products will enable the determination of the magnetic connectivity of the spacecraft to the planet and its plasma environment. In addition, M-EAS has inherited the burst-mode capability from Solar Orbiter's EAS.  It will also measure pitch-angle distributions of negative ions, which are  products of certain photo-chemical processes in the Martian ionosphere \citep{kopp00,molina02}. The \textbf{Mars - Ion and Neutral Energy Analyser (M-INEA)} is a novel energy spectrograph and mass spectrometer designed to sample the low-energy range of the neutral particle distribution at energies below 10\,eV with a high energy resolution of about 0.1\,eV. It builds on heritage from BepiColombo, Solar Orbiter, and PRELUDE-SAT.  The instrument will capture the energy spectrum and the mass of the incoming particles simultaneously. For the first time, M-INEA will provide us with direct measurements of the neutral atmospheric escape rate of the Martian atmosphere.  The \textbf{Solar Particles at Mars (SP@M)} sensor consists of four ion and four electron solid-state detector units to measure solar energetic particles and suprathermal particle populations near Mars. Its design follows heritage from MAVEN and Taranis. SP@M will record ions in the energy range from 30\,keV to 10\,MeV and electrons in the range from 30\,keV to 1\,MeV. These measurements will be crucial for the understanding of atmospheric escape, aurorae induced by solar energetic particles, and radar blackouts in the Martian ionosphere under active conditions.
\item The \textbf{Combined Magnetic and Plasma Sensor Suite (COMPASS)} is also a suite of instruments on M-MATISSE. Its four instruments will measure the electromagnetic fields and certain particle properties in the Martian environment. The \textbf{Fluxgate Magnetometer (MAG)} will measure the vector magnetic field at the location of the spacecraft with an instrument design that builds on heritage from JUICE and Comet Interceptor. The \textbf{Dual Langmuir Probes (LP)} will characterise the electron density and temperature, the spacecraft potential, and one component of the low-frequency electric field. The LP measurements of the electron density and temperature will be independent of the electron measurements taken by M-EAS and thus serve for calibration purposes with M-EAS. The \textbf{Mutual Impedance eXperiment (MIX)} will use the mutual-impedance technique to measure the electron density and temperature, as well as the high-frequency electric field. This instrument design is inherited from similar instruments on Rosetta, BepiColombo, JUICE, and Comet Interceptor. The mutual-impedance technique involves the measurement of the frequency response of the coupling impedance between two dipoles.  The current-voltage characteristics of the two electrodes shows a resonance peak and interference patterns from which the plasma quantities are then extracted. The \textbf{3D Velocity of Ions (3DVI)} instrument consists of an ion drift meter combined with a retarding potential analyser. Through the measurement of the incoming particle currents, it will determine the three-dimensional velocity vector and the temperature of thermal ions in the Martian ionosphere where particle densities are high. 
\item The \textbf{Mars - Mass Spectrum Analyser (M-MSA)} is a top-hat electrostatic ion analyser with electrostatic aperture deflectors, combined with a time-of-flight chamber to determine the mass-per-charge ratio of the incoming particles. It builds on heritage from BepiColombo and the Martian Moons eXploration (MMX) mission. M-MSA will resolve energy-per-charge from 2\,eV/e to 38\,keV/e and particle mass $m$ with a resolution of $m/\Delta m\approx 40$. 
\item The \textbf{Mars - Solar Spectral Irradiance Monitor (M-SoSpIM)} is a monitor for extreme ultraviolet and x-ray radiation with heritage from Solar-C and MAVEN. M-SoSpIM will record the solar irradiance in its energy channels every second and at sub-second cadence during solar occultations. It will provide three channels: soft x-rays to observe radiation from the upper solar corona and emitted during solar flares, extreme ultraviolet to monitor the solar corona as the source region of space-weather events, and Lyman-$\alpha$ radiation to link these observations to the lower solar atmosphere. 
\item The \textbf{M-MATISSE Crosslink Experiment (MaCro)} is an inter-satellite radio experiment to probe the Martian ionosphere and atmosphere in occultation between both spacecraft. Mars Express, Venus Express, and EnVision serve as heritage for this instrument design. MaCro will operate transceivers with ultrastable oscillators at two frequencies. It will measure the shift of the carrier frequencies and the attenuation of the signal caused by the passing of the radio waves through the ionosphere and atmosphere. From these measurements of the inter-spacecraft link, MaCro will determine vertical profiles of the electron density as well as the lateral and vertical electron contents in the ionosphere. It will also record temperature, pressure, and the neutral particles' number density profiles in the neutral atmosphere. 
\item The \textbf{Mars - Aurora and Dust Camera (M-AC)} is an auroral camera that will detect the green aurora on the Martian nightside, and observe the atmospheric dust opacity and ice clouds in the lower atmosphere on the Martian dayside. It builds on heritage from a cubesat mission in low-Earth orbit and IMAP. The observation of aurora is crucial for the understanding of the energy deposition from space into the Martian atmosphere. For its dual-use capability, M-AC will require a very large dynamic range to measure both at night and during day.
\end{itemize}

The main challenges for the M-MATISSE mission result from the dual-spacecraft character of the mission. The correlation of upstream properties with the response of the Martian system requires novel analysis methods. In addition, many of the mission's in-situ instruments require a wide dynamic range, allowing them to sample conditions ranging from the very dilute plasma of the upstream solar wind to the dense ionospheric plasma near the planet.

M-MATISSE will be the next generation of Mars missions, extending the Mars Atmosphere and Volatile EvolutioN (MAVEN) mission \citep{jakosky15}, which was launched in 2013, and the Escape and Plasma Acceleration and Dynamics Explorers (ESCAPADE) mission, which was launched in 2025.  MAVEN's central science goal is the study of atmospheric loss at Mars, while M-MATISSE will focus on the magnetosphere-ionosphere-thermosphere coupling. M-MATISSE will provide the understanding of the full atmospheric column from the surface to space and thus target the full chain of processes leading to atmospheric escape from the neutral reservoir into open space. ESCAPADE is a low-cost two-spacecraft mission that carries a miniaturised payload suite including a magnetometer, an electrostatic analyser, and a Langmuir probe, all with limited power, performance, and telemetry capabilities. ESCAPADE  provides a global overview of the flow of energy from the solar wind through Mars' hybrid magnetosphere and the resulting driving of escape processes. M-MATISSE, however, will not just measure all observables targeted by ESCAPADE at higher resolution. It will additionally capture critical parameters that ESCAPADE cannot observe: solar extreme ultraviolet flux, neutral densities and fluxes, electric fields, plasma temperatures, densities and temperatures in the lower atmosphere, and energetic particles.

\subsection{Debye}

Debye is a mission concept that was submitted to the first and second calls for ESA F-class science missions. It is designed as the first dedicated electron-astrophysics mission with the goal of exploring the physics of electron heating in astrophysical  plasmas \citep{verscharen22}. Electron-kinetic scales are the smallest characteristic scales associated with collective plasma processes. Therefore, the measurement of electron-scale processes requires high spatial and temporal resolution \citep{alexandrova09,roberts17}. The Debye mission will uncover the nature of electric-field and magnetic-field fluctuations at electron scales, discover how electron-scale fluctuations heat plasma electrons, determine how energy is partitioned between ions and electrons, and investigate how electron thermalisation varies with the overall plasma conditions.

The first iteration of the Debye mission proposal foresees one main spacecraft and up to three deployable spacecraft, which will provide simultaneous multi-point measurements of the magnetic-field fluctuations in the solar wind. The spacecraft separation will be of order a few 100\,m to tens of km. These scales cover electron scales as well as the transition from ion scales into electron scales. 

The Debye main spacecraft will carry the following instruments:
\begin{itemize}
\item The \textbf{Thermal Electron Analyser (TEA)} is a top-hat electrostatic analyser with aperture deflection system for the measurement of thermal and suprathermal electrons. It builds on heritage from Solar Orbiter's EAS, but with a larger geometric factor to enable faster measurements of the electron pitch-angle distributions. To resolve the relevant electron scales, TEA will record one electron pitch-angle distribution every 50\,ms. The TEA measurements will be crucial for the identification of electron-kinetic dissipation mechanisms in the plasma.
\item The \textbf{Proton Electrostatic Analyser (PEA)} is a top-hat electrostatic analyser with aperture deflection system for the measurement of thermal protons, building on heritage from Solar Orbiter's PAS design.  PEA will provide reliable measurements of the proton moments, which are required to disentangle spatial and temporal fluctuations in the plasma  based on the knowledge of the velocity with which these structures are convected across the spacecraft formation \citep{sahraoui10,sahraoui11}. Moreover, it will determine the plasma context (e.g., in terms of the plasma-$\beta$) and quantify the partition of energy transfer into proton heating.
\item The \textbf{Fluxgate Magnetometer (FGM)}  will measure the background magnetic field as well as fluctuations in the magnetic field at proton scales and larger scales. It builds on heritage from Solar Orbiter and JUICE. FGM will provide context for the local plasma conditions, including the overall turbulence environment, in which the electron-kinetic processes operate.
\item The \textbf{Electric Field Instrument (EFI)} consists of four spherical probes located on the tips of 20-m-long wire booms. The instrument builds on heritage from BepiColombo. EFI will measure the electric field in the frequency range between 0 and 100\,kHz. The measurement of the electric field will provide reliable estimates of the phase velocity of electric-field structures up to 10,000\,km/s. Moreover, EFI will provide a measurement of the spacecraft floating potential and an independent measurement of the electron density for cross-calibration with TEA.
\end{itemize}

The main spacecraft as well as all deployable spacecraft will each carry a \textbf{Search Coil Magnetometer (SCM)} to measure high-frequency fluctuations in the vector magnetic field. The SCM design carries heritage from Solar Orbiter and JUICE.  The multi-point measurement of high-frequency fluctuations in the magnetic field will be crucial for the identification of electromagnetic electron-scale fluctuations in the plasma. Through the combination of simultaneous measurements of the magnetic-field fluctuations at multiple points with single-point measurements of the electron velocity distribution function, Debye will characterise the kinetic physics of electron-scale processes in collisionless plasmas. The SCM and the EFI instruments will be supported by a \textbf{Wave Electronics Box (WEB)}. The TEA and PEA instruments will be supported by a \textbf{Particle Data Processor (PDP)}.

The main challenges for the Debye mission result from the smallness of electron-kinetic scales. Their measurement requires small baselines between the involved spacecraft, which is a technological challenge in terms of spacecraft operations. Furthermore, electron-scale observations require high time resolution in all measurements and high velocity-space resolution in the measurement of the electron velocity distribution function. Debye has undergone multiple iterations from the described design, responding to scientific and programmatic requirements defined by the given calls for mission proposals. Nevertheless, the design and flight of a dedicated electron-astrophysics mission is an exciting scientific opportunity, as it relates electron-scale plasma physics with the plasma physics of some of the largest objects in the Universe.

\section{Conclusions}

Space plasmas offer unique opportunities to measure fundamental plasma processes in situ. With modern instrumentation, such as electrostatic analysers, we measure the velocity distribution functions of ions and electrons in these collisionless plasma systems to a very high level of detail. A major advantage of these space-plasma measurements is their micro-invasive nature, in the sense that perturbations to the plasma due to our diagnostic devices are often negligible. This advantage of in-situ measurements taken in space makes them a powerful tool for the understanding of plasma physics in general.

A fleet of successful space missions with instrumentation that observes both the plasma particles and the electromagnetic fields simultaneously is currently operating in the heliosphere, and the next generation of new space missions with these capabilities is currently in development. For the benefit of the understanding of fundamental plasma physics, it would be worthwhile to explore in more detail the synergies between the research communities working on diagnostic methods in space plasma physics and on diagnostic methods in laboratory plasma physics.
Active plasma experiments in space are another avenue for fruitful interactions between researchers working in space and laboratory plasma physics \citep{winckler92,gekelman95,koepke08,pongratz18,borovsky19,haerendel19}. 

There is a great potential for the space and the laboratory plasma physics communities to learn from each other and to achieve joint progress in our understanding of plasma physics. In fact, it used to be more common for plasma physicists to work across the discipline boundaries in the past. As the field has evolved, however, the differences in nomenclature, scientific targets, training and education programmes, and research methodologies have grown. Nevertheless, there are strong benefits of synergistic and interdisciplinary research in the domain of fundamental plasma physics that are worth being explored in the future.

\ack
I am very grateful for the invitation to present this review at the 6\textsuperscript{th} European Conference on Plasma Diagnostics (ECPD 2025) in Prague. I appreciate helpful discussions with Chris Owen, Andrew Fazakerley, and George Nicolaou about the intricacies of electrostatic analysers, and I am grateful to Kris Klein, Beatriz Sanchez-Cano, and Rob Wicks for our discussions about HelioSwarm,  M-MATISSE, and Debye. I express my gratitude to MSSL's engineering and technology teams involved in the design and development of Solar Orbiter's SWA, Vigil's PLA, M-MATISSE's M-EAS, and HelioSwarm's iESA, as well as the wider international mission teams. This work is supported by STFC's Consolidated Grant ST/W001004/1 and UKSA funding for M-MATISSE, administered through STFC grant ST/Z000742/1.

\newcommand{\newblock}{}
\bibliographystyle{jphysicsB}
\bibliography{ECPD_review_production}

@ARTICLE{verscharen19,
       author = {{Verscharen}, Daniel and {Klein}, Kristopher G. and {Maruca}, Bennett A.},
        title = "{The multi-scale nature of the solar wind}",
      journal = {Living Rev.~Solar Phys.},
         year = 2019,
        month = dec,
       volume = {16},
       number = {1},
          eid = {5},
        pages = {5},
          doi = {10.1007/s41116-019-0021-0},
archivePrefix = {arXiv},
       eprint = {1902.03448},
 primaryClass = {physics.space-ph},
       adsurl = {https://ui.adsabs.harvard.edu/abs/2019LRSP...16....5V}
}

@BOOK{bryant99,
       author = {{Bryant}, D.~A.},
        title = "{Electron acceleration in the aurora and beyond}",
         year = 1999,
         publisher = {IOP Publishing Ltd, CRC Press, Bristol, Philadelphia},
         doi = {10.1201/9781482268539},
       adsurl = {https://ui.adsabs.harvard.edu/abs/1999eaab.book.....B}
}

@ARTICLE{verscharen21,
       author = {{Verscharen}, Daniel and {Wicks}, Robert T. and {Branduardi-Raymont}, Graziella and {Erd{\'e}lyi}, Robertus and {Frontera}, Filippo and {G{\"o}tz}, Charlotte and {Guidorzi}, Cristiano and {Lebouteiller}, Vianney and {Matthews}, Sarah A. and {Nicastro}, Fabrizio and {Rae}, Iain Jonathan and {Retin{\`o}}, Alessandro and {Simionescu}, Aurora and {Soffitta}, Paolo and {Uttley}, Phil and {Wimmer-Schweingruber}, Robert F.},
        title = "{The Plasma Universe: A Coherent Science Theme for Voyage 2050}",
      journal = {Frontiers in Astronomy and Space Sciences},
         year = 2021,
        month = apr,
       volume = {8},
          eid = {30},
        pages = {30},
          doi = {10.3389/fspas.2021.651070},
archivePrefix = {arXiv},
       eprint = {2104.07983},
 primaryClass = {physics.plasm-ph},
       adsurl = {https://ui.adsabs.harvard.edu/abs/2021FrASS...8...30V}
}

@ARTICLE{pfaff98,
       author = {{Pfaff}, Robert F. and {Borovsky}, Joseph E. and {Young}, David T.},
        title = "{Measurement Techniques in Space Plasmas -- Particles}",
      journal = {Geophysical Monograph Series},
         year = 1998,
        month = jan,
       volume = {102},
          doi = {10.1029/GM102},
       adsurl = {https://ui.adsabs.harvard.edu/abs/1998GMS...102.....P}
}

@ARTICLE{pfaff98b,
       author = {{Pfaff}, Robert F. and {Borovsky}, Joseph E. and {Young}, David T.},
        title = "{Measurement Techniques in Space Plasmas -- Fields}",
      journal = {Geophysical Monograph Series},
         year = 1998,
        month = jan,
       volume = {103},
          doi = {10.1029/GM103},
       adsurl = {https://ui.adsabs.harvard.edu/abs/1998GMS...103.....P}
}

@ARTICLE{marsch06,
       author = {{Marsch}, Eckart},
        title = "{Kinetic Physics of the Solar Corona and Solar Wind}",
      journal = {Living Reviews in Solar Physics},
         year = 2006,
        month = dec,
       volume = {3},
       number = {1},
          eid = {1},
        pages = {1},
          doi = {10.12942/lrsp-2006-1},
       adsurl = {https://ui.adsabs.harvard.edu/abs/2006LRSP....3....1M}
}

@ARTICLE{wilson18,
       author = {{Wilson}, III, Lynn B. and {Stevens}, Michael L. and {Kasper}, Justin C. and {Klein}, Kristopher G. and {Maruca}, Bennett A. and {Bale}, Stuart D. and {Bowen}, Trevor A. and {Pulupa}, Marc P. and {Salem}, Chadi S.},
        title = "{The Statistical Properties of Solar Wind Temperature Parameters Near 1 au}",
      journal = {\apjs},
         year = 2018,
        month = jun,
       volume = {236},
       number = {2},
          eid = {41},
        pages = {41},
          doi = {10.3847/1538-4365/aab71c},
archivePrefix = {arXiv},
       eprint = {1802.08585},
 primaryClass = {physics.plasm-ph},
       adsurl = {https://ui.adsabs.harvard.edu/abs/2018ApJS..236...41W}
}

@ARTICLE{mostafavi24,
       author = {{Mostafavi}, P. and {Allen}, R.~C. and {Jagarlamudi}, V.~K. and {Bourouaine}, S. and {Badman}, S.~T. and {Ho}, G.~C. and {Raouafi}, N.~E. and {Hill}, M.~E. and {Verniero}, J.~L. and {Larson}, D.~E. and {Kasper}, J.~C. and {Bale}, S.~D.},
        title = "{Parker Solar Probe observations of collisional effects on thermalizing the young solar wind}",
      journal = {\aap},
         year = 2024,
        month = feb,
       volume = {682},
          eid = {A152},
        pages = {A152},
          doi = {10.1051/0004-6361/202347134},
       adsurl = {https://ui.adsabs.harvard.edu/abs/2024A&A...682A.152M}
}

@ARTICLE{johnson24,
       author = {{Johnson}, E. and {Maruca}, B.~A. and {McManus}, M. and {Stevens}, M. and {Klein}, K.~G. and {Mostafavi}, P.},
        title = "{Application of collisional analysis to the differential velocity of solar wind ions}",
      journal = {Front.~Astron.~Space Sci.},
         year = 2024,
        month = jan,
       volume = {10},
          eid = {1284913},
        pages = {1284913},
          doi = {10.3389/fspas.2023.1284913},
       adsurl = {https://ui.adsabs.harvard.edu/abs/2024FrASS..1084913J}
}

@ARTICLE{bercic21c,
       author = {{Ber{\v{c}}i{\v{c}}}, L. and {Landi}, S. and {Maksimovi{\'c}}, M.},
        title = "{The Interplay Between Ambipolar Electric Field and Coulomb Collisions in the Solar Wind Acceleration Region}",
      journal = {J.~Geophys.~Res.~(Space Phys.)},
         year = 2021,
        month = mar,
       volume = {126},
       number = {3},
          eid = {e28864},
        pages = {e28864},
          doi = {10.1029/2020JA028864},
       adsurl = {https://ui.adsabs.harvard.edu/abs/2021JGRA..12628864B}
}

@ARTICLE{heidrich20,
       author = {{Heidrich-Meisner}, Verena and {Berger}, Lars and {Wimmer-Schweingruber}, Robert F.},
        title = "{Proton-proton collisional age to order solar wind types}",
      journal = {\aap},
         year = 2020,
        month = apr,
       volume = {636},
          eid = {A103},
        pages = {A103},
          doi = {10.1051/0004-6361/201937378},
archivePrefix = {arXiv},
       eprint = {2003.10851},
 primaryClass = {astro-ph.SR},
       adsurl = {https://ui.adsabs.harvard.edu/abs/2020A&A...636A.103H}
}

@ARTICLE{bale13,
       author = {{Bale}, S.~D. and {Pulupa}, M. and {Salem}, C. and {Chen}, C.~H.~K. and {Quataert}, E.},
        title = "{Electron Heat Conduction in the Solar Wind: Transition from Spitzer-H{\"a}rm to the Collisionless Limit}",
      journal = {\apjl},
         year = 2013,
        month = jun,
       volume = {769},
       number = {2},
          eid = {L22},
        pages = {L22},
          doi = {10.1088/2041-8205/769/2/L22},
archivePrefix = {arXiv},
       eprint = {1303.0932},
 primaryClass = {astro-ph.SR},
       adsurl = {https://ui.adsabs.harvard.edu/abs/2013ApJ...769L..22B}
}

@ARTICLE{johnstone85,
       author = {{Johnstone}, A.~D. and {Kellock}, S.~J. and {Coates}, A.~J. and {Smith}, M.~F. and {Booker}, T. and {Winningham}, J.~D.},
        title = "{A space-borne plasma analyser for three-dimensional measurements of the velocity distribution}",
      journal = {IEEE Trans.~Nucl.~Sci.},
         year = 1985,
        month = feb,
       volume = {32},
        pages = {139-144},
          doi = {10.1109/TNS.1985.4336809},
       adsurl = {https://ui.adsabs.harvard.edu/abs/1985ITNS...32..139J}
}

@ARTICLE{zhang24,
       author = {{Zhang}, H. and {Verscharen}, D. and {Nicolaou}, G.},
        title = "{The Impact of Non-Equilibrium Plasma Distributions on Solar Wind Measurements by Vigil's Plasma Analyser}",
      journal = {Space Weather},
         year = 2024,
        month = feb,
       volume = {22},
       number = {2},
          eid = {e2023SW003671},
        pages = {e2023SW003671},
          doi = {10.1029/2023SW003671},
archivePrefix = {arXiv},
       eprint = {2402.04694},
 primaryClass = {physics.space-ph},
       adsurl = {https://ui.adsabs.harvard.edu/abs/2024SpWea..2203671Z}
}

@ARTICLE{fazakerley98,
       author = {{Fazakerley}, Andrew N. and {Schwartz}, Steven J. and {Paschmann}, G{\"o}tz},
        title = "{Measurement of Plasma Velocity Distributions}",
      journal = {ISSI Scientific Reports Series},
         year = 1998,
        month = jan,
       volume = {1},
        pages = {91-124},
       adsurl = {https://ui.adsabs.harvard.edu/abs/1998ISSIR...1...91F}
}

@ARTICLE{demarco23,
       author = {{De Marco}, R. and {Bruno}, R. and {Jagarlamudi}, V. Krishna and {D'Amicis}, R. and {Marcucci}, M.~F. and {Fortunato}, V. and {Perrone}, D. and {Telloni}, D. and {Owen}, C.~J. and {Louarn}, P. and {Fedorov}, A. and {Livi}, S. and {Horbury}, T.},
        title = "{Innovative technique for separating proton core, proton beam, and alpha particles in solar wind 3D velocity distribution functions}",
      journal = {\aap},
         year = 2023,
        month = jan,
       volume = {669},
          eid = {A108},
        pages = {A108},
          doi = {10.1051/0004-6361/202243719},
       adsurl = {https://ui.adsabs.harvard.edu/abs/2023A&A...669A.108D}
}

@ARTICLE{cho23,
       author = {{Cho}, Kyung-Suk and {Hwang}, Junga and {Han}, Jeong-Yeol and {Choi}, Seong-Hwan and {Park}, Sung-Hong and {Lim}, Eun-Kyung and {Kim}, Rok-Soon and {Seough}, Jungjoon and {Sohn}, Jong-Dae and {Song}, Donguk and {Kwak}, Jae-Young and {Miyashita}, Yukinaga and {Baek}, Ji-Hye and {Lee}, Jaejin and {Lee}, Jinsung and {Ryu}, Kwangsun and {Seon}, Jongho and {Jin}, Ho and {Ye}, Sung-Jun and {Moon}, Yong-Jae and {Lee}, Dae-Young and {Yoon}, Peter H. and {Hoang}, Thiem and {Sterken}, Veerle and {Joshi}, Bhuwan and {Lee}, Chang-Han and {Jang}, Jongjin and {Doh}, Jae-Hwee and {Kim}, Hwayeong and {Park}, Hyeon-Jeong and {Gopalswamy}, Natchimuthuk and {Elsayed}, Talaat and {Lee}, John},
        title = "{Opening New Horizons with the L4 Mission: Vision and Plan}",
      journal = {J.Korean Astron.~Soc.},
         year = 2023,
        month = nov,
       volume = {56},
        pages = {263-275},
          doi = {10.5303/JKAS.2023.56.2.263},
       adsurl = {https://ui.adsabs.harvard.edu/abs/2023JKAS...56..263C}
}

@ARTICLE{bemporad21,
       author = {{Bemporad}, A.},
        title = "{Possible advantages of a twin spacecraft Heliospheric mission at the Sun-Earth Lagrangian points L4 and L5}",
      journal = {Front.~Astron.~Space Sci.},
         year = 2021,
        month = mar,
       volume = {8},
          eid = {11},
        pages = {11},
          doi = {10.3389/fspas.2021.627576},
       adsurl = {https://ui.adsabs.harvard.edu/abs/2021FrASS...8...11B}
}

@ARTICLE{pongratz18,
       author = {{Pongratz}, Morris B.},
        title = "{History of Los Alamos Participation in Active Experiments in Space}",
      journal = {Front.~Phys.},
         year = 2018,
        month = dec,
       volume = {6},
          eid = {144},
        pages = {144},
          doi = {10.3389/fphy.2018.00144},
       adsurl = {https://ui.adsabs.harvard.edu/abs/2018FrP.....6..144P}
}

@ARTICLE{haerendel19,
       author = {{Haerendel}, Gerhard},
        title = "{Experiments With Plasmas Artificially Injected Into Near-Earth Space}",
      journal = {Front.~Astron.~Space Sci.},
         year = 2019,
        month = apr,
       volume = {6},
          eid = {29},
        pages = {29},
          doi = {10.3389/fspas.2019.00029},
       adsurl = {https://ui.adsabs.harvard.edu/abs/2019FrASS...6...29H}
}

@ARTICLE{winckler92,
       author = {{Winckler}, John R.},
        title = "{Controlled experiments in the earth's magnetosphere with artifical electron beams}",
      journal = {Rev.~Modern Phys.},
         year = 1992,
        month = jul,
       volume = {64},
       number = {3},
        pages = {859-871},
          doi = {10.1103/RevModPhys.64.859},
       adsurl = {https://ui.adsabs.harvard.edu/abs/1992RvMP...64..859W}
}

@ARTICLE{koepke08,
       author = {{Koepke}, M.~E.},
        title = "{Interrelated laboratory and space plasma experiments}",
      journal = {Rev.~Geophys.},
         year = 2008,
        month = sep,
       volume = {46},
       number = {3},
          eid = {RG3001},
        pages = {RG3001},
          doi = {10.1029/2005RG000168},
       adsurl = {https://ui.adsabs.harvard.edu/abs/2008RvGeo..46.3001K}
}

@ARTICLE{borovsky19,
       author = {{Borovsky}, Joseph E. and {Delzanno}, Gian Luca},
        title = "{Active Experiments in Space: The Future}",
      journal = {Front.~Astron.~Space Sci.},
         year = 2019,
        month = may,
       volume = {6},
          eid = {31},
        pages = {31},
          doi = {10.3389/fspas.2019.00031},
       adsurl = {https://ui.adsabs.harvard.edu/abs/2019FrASS...6...31B}
}

@ARTICLE{gekelman95,
       author = {{Gekelman}, Walter},
        title = "{Active and laboratory experiments in space plasma physics}",
      journal = {Surv.~Geophys.},
         year = 1995,
        month = may,
       volume = {16},
       number = {3},
        pages = {457-485},
          doi = {10.1007/BF01044576},
       adsurl = {https://ui.adsabs.harvard.edu/abs/1995SGeo...16..457G}
}

@ARTICLE{mutel07,
       author = {{Mutel}, R.~L. and {Peterson}, W.~M. and {Jaeger}, T.~R. and {Scudder}, J.~D.},
        title = "{Dependence of cyclotron maser instability growth rates on electron velocity distributions and perturbation by solitary waves}",
      journal = {\jgr},
         year = 2007,
        month = jul,
       volume = {112},
       number = {A7},
          eid = {A07211},
        pages = {A07211},
          doi = {10.1029/2007JA012442},
archivePrefix = {arXiv},
       eprint = {0705.0877},
 primaryClass = {astro-ph},
       adsurl = {https://ui.adsabs.harvard.edu/abs/2007JGRA..112.7211M}
}

@ARTICLE{mcfadden99,
       author = {{McFadden}, J.~P. and {Carlson}, C.~W. and {Ergun}, R.~E.},
        title = "{Microstructure of the auroral acceleration region as observed by FAST}",
      journal = {\jgr},
         year = 1999,
        month = jul,
       volume = {104},
       number = {A7},
        pages = {14453-14480},
          doi = {10.1029/1998JA900167},
       adsurl = {https://ui.adsabs.harvard.edu/abs/1999JGR...10414453M}
}

@ARTICLE{carlson98,
       author = {{Carlson}, C.~W. and {Pfaff}, R.~F. and {Watzin}, J.~G.},
        title = "{The Fast Auroral SnapshoT (FAST) Mission}",
      journal = {\grl},
         year = 1998,
        month = jun,
       volume = {25},
       number = {12},
        pages = {2013-2016},
          doi = {10.1029/98GL01592},
       adsurl = {https://ui.adsabs.harvard.edu/abs/1998GeoRL..25.2013C}
}

@ARTICLE{vaivads95,
       author = {{Vaivads}, A. and {R{\"o}nnmark}, K. and {Andr{\'e}}, M.},
        title = "{Generation of ion acoustic waves by fan instability}",
      journal = {\jgr},
         year = 1995,
        month = oct,
       volume = {100},
       number = {A10},
        pages = {19435-19440},
          doi = {10.1029/95JA01629},
       adsurl = {https://ui.adsabs.harvard.edu/abs/1995JGR...10019435V}
}

@ARTICLE{dendy93,
       author = {{Dendy}, R.~O. and {Lashmore-Davies}, C.~N. and {Kam}, K.~F.},
        title = "{The magnetoacoustic cyclotron instability of an extended shell distribution of energetic ions}",
      journal = {Phys.~Fluids B},
         year = 1993,
        month = jul,
       volume = {5},
       number = {7},
        pages = {1937-1944},
          doi = {10.1063/1.860781},
       adsurl = {https://ui.adsabs.harvard.edu/abs/1993PhFlB...5.1937D}
}

@ARTICLE{riedler97,
       author = {{Riedler}, W. and {Torkar}, K. and {Rudenauer}, F. and {Fehringer}, M. and {Pedersen}, A. and {Schmidt}, R. and {Grard}, R.~J.~L. and {Arends}, H. and {Narheim}, B.~T. and {Troim}, J. and et al.},
        title = "{Active Spacecraft Potential Control}",
      journal = {\ssr},
         year = 1997,
        month = jan,
       volume = {79},
        pages = {271-302},
          doi = {10.1023/A:1004921614592},
       adsurl = {https://ui.adsabs.harvard.edu/abs/1997SSRv...79..271R}
}

@ARTICLE{parail80,
       author = {{Parail}, V.~V. and {Pogutse}, O.~P.},
        title = "{Fan instability and anomalous ion heating}",
      journal = {ZhETF Pisma Redaktsiiu},
         year = 1980,
        month = feb,
       volume = {31},
        pages = {165-168},
       adsurl = {https://ui.adsabs.harvard.edu/abs/1980ZhPmR..31..165P}
}

@ARTICLE{torkar16,
       author = {{Torkar}, K. and {Nakamura}, R. and {Tajmar}, M. and {Scharlemann}, C. and {Jeszenszky}, H. and {Laky}, G. and {Fremuth}, G. and {Escoubet}, C.~P. and {Svenes}, K.},
        title = "{Active Spacecraft Potential Control Investigation}",
      journal = {\ssr},
         year = 2016,
        month = mar,
       volume = {199},
       number = {1-4},
        pages = {515-544},
          doi = {10.1007/s11214-014-0049-3},
       adsurl = {https://ui.adsabs.harvard.edu/abs/2016SSRv..199..515T}
}

@ARTICLE{stverak26,
       author = {{{\v{S}}tver{\'a}k}, {\v{S}}. and {Her{\v{c}}{\'\i}k}, D. and {Hellinger}, P. and {Pop{\v{d}}akunik}, M. and {Lewis}, G.~R. and {Nicolaou}, G. and {Owen}, C.~J. and {Khotyaintsev}, Yu. V. and {Maksimovic}, M.},
        title = "{Modelling spacecraft-emitted electrons measured by SWA-EAS experiment on board Solar Orbiter mission}",
      journal = {arXiv e-prints},
         year = 2026,
        month = jan,
          eid = {arXiv:2601.03818},
        pages = {arXiv:2601.03818},
          doi = {10.48550/arXiv.2601.03818},
archivePrefix = {arXiv},
       eprint = {2601.03818},
 primaryClass = {astro-ph.IM},
       adsurl = {https://ui.adsabs.harvard.edu/abs/2026arXiv260103818S}
}

@ARTICLE{wilson23,
       author = {{Wilson}, III, Lynn B. and {Salem}, Chadi S. and {Bonnell}, John W.},
        title = "{Spacecraft Floating Potential Measurements for the Wind Spacecraft}",
      journal = {\apjs},
         year = 2023,
        month = dec,
       volume = {269},
       number = {2},
          eid = {52},
        pages = {52},
          doi = {10.3847/1538-4365/ad0633},
archivePrefix = {arXiv},
       eprint = {2309.11626},
 primaryClass = {physics.space-ph},
       adsurl = {https://ui.adsabs.harvard.edu/abs/2023ApJS..269...52W}
}

@ARTICLE{scudder00,
       author = {{Scudder}, J.~D. and {Cao}, Xuejun and {Mozer}, F.~S.},
        title = "{Photoemission current-spacecraft voltage relation: Key to routine, quantitative low-energy plasma measurements}",
      journal = {\jgr},
         year = 2000,
        month = sep,
       volume = {105},
       number = {A9},
        pages = {21,281-21,294},
          doi = {10.1029/1999JA900423},
       adsurl = {https://ui.adsabs.harvard.edu/abs/2000JGR...10521281S}
}

@ARTICLE{scime94,
       author = {{Scime}, Earl E. and {Phillips}, John L. and {Bame}, Samuel J.},
        title = "{Effects of spacecraft potential on three-dimensional electron measurements in the solar wind}",
      journal = {\jgr},
         year = 1994,
        month = aug,
       volume = {99},
       number = {A8},
        pages = {14769-14776},
          doi = {10.1029/94JA00489},
       adsurl = {https://ui.adsabs.harvard.edu/abs/1994JGR....9914769S}
}

@ARTICLE{salem01,
       author = {{Salem}, C. and {Bosqued}, J.-M. and {Larson}, D.~E. and {Mangeney}, A. and {Maksimovic}, M. and {Perche}, C. and {Lin}, R.~P. and {Bougeret}, J.-L.},
        title = "{Determination of accurate solar wind electron parameters using particle detectors and radio wave receivers}",
      journal = {\jgr},
         year = 2001,
        month = oct,
       volume = {106},
       number = {A10},
        pages = {21701-21717},
          doi = {10.1029/2001JA900031},
       adsurl = {https://ui.adsabs.harvard.edu/abs/2001JGR...10621701S}
}

@ARTICLE{bergman20,
       author = {{Bergman}, Sofia and {Stenberg Wieser}, Gabriella and {Wieser}, Martin and {Johansson}, Fredrik Leffe and {Eriksson}, Anders},
        title = "{The Influence of Spacecraft Charging on Low-Energy Ion Measurements Made by RPC-ICA on Rosetta}",
      journal = {J.~Geophys.~Res.~(Space Phys.)},
         year = 2020,
        month = jan,
       volume = {125},
       number = {1},
          eid = {e27478},
        pages = {e27478},
          doi = {10.1029/2019JA027478},
       adsurl = {https://ui.adsabs.harvard.edu/abs/2020JGRA..12527478B}
}

@ARTICLE{song97,
       author = {{Song}, P. and {Zhang}, X.~X. and {Paschmann}, G.},
        title = "{Uncertainties in plasma measurements: effects of lower cutoff energy and spacecraft charge}",
      journal = {\planss},
         year = 1997,
        month = feb,
       volume = {45},
       number = {2},
        pages = {255-267},
          doi = {10.1016/S0032-0633(96)00087-6},
       adsurl = {https://ui.adsabs.harvard.edu/abs/1997P&SS...45..255S}
}

@ARTICLE{berry81,
       author = {{Berry Garrett}, Henry},
        title = "{The Charging of Spacecraft Surfaces (Paper 1R1000)}",
      journal = {Rev.~Geophys.~Space Phys.},
         year = 1981,
        month = nov,
       volume = {19},
        pages = {577},
          doi = {10.1029/RG019i004p00577},
       adsurl = {https://ui.adsabs.harvard.edu/abs/1981RvGSP..19..577B}
}

@ARTICLE{whipple81,
       author = {{Whipple}, E.~C.},
        title = "{Potentials of surfaces in space}",
      journal = {Rep.~Prog.~Phys.},
         year = 1981,
        month = nov,
       volume = {44},
       number = {11},
        pages = {1197-1250},
          doi = {10.1088/0034-4885/44/11/002},
       adsurl = {https://ui.adsabs.harvard.edu/abs/1981RPPh...44.1197W}
}

@ARTICLE{sahraoui11,
       author = {{Sahraoui}, Fouad and {Goldstein}, Melvyn L. and {Belmont}, G{\'e}rard and {Roux}, Alain and {Rezeau}, Laurence and {Canu}, Patrick and {Robert}, Patrick and {Cornilleau-Wehrlin}, Nicole and {Le Contel}, Olivier and {Dudok De Wit}, Thierry and {Pin{\c{c}}on}, Jean-louis and {Kiyani}, Khurom},
        title = "{Multi-spacecraft investigation of space turbulence: Lessons from Cluster and input to the Cross-Scale mission}",
      journal = {\planss},
         year = 2011,
        month = may,
       volume = {59},
       number = {7},
        pages = {585-591},
          doi = {10.1016/j.pss.2010.06.001},
       adsurl = {https://ui.adsabs.harvard.edu/abs/2011P&SS...59..585S}
}

@ARTICLE{lee24,
       author = {{Lee}, Dae-Young and {Kim}, Rok-Soon and {Choi}, Kyung-Eun and {Seough}, Jungjoon and {Hwang}, Junga and {Choi}, Dooyoung and {Yoo}, Ji-Hyeon and {Lee}, Seunguk and {Noh}, Sung Jun and {Seon}, Jongho and {Cho}, Kyung-Suk and {Ryu}, Kwangsun and {Kim}, Khan-Hyuk and {Sohn}, Jong-Dae and {Kwak}, Jae-Young and {Yoon}, Peter H.},
        title = "{Long-Term Science Goals with In Situ Observations at the Sun-Earth Lagrange Point L4}",
      journal = {J.~Astron.~Space Sci.},
         year = 2024,
        month = mar,
       volume = {41},
       number = {1},
        pages = {1-15},
          doi = {10.5140/JASS.2024.41.1.1},
       adsurl = {https://ui.adsabs.harvard.edu/abs/2024JASS...41....1L}
}

@ARTICLE{moon24,
       author = {{Moon}, Yong-Jae and {Cho}, Kyung-Suk and {Park}, Sung-Hong and {Lim}, Eun-Kyung and {Kim}, Roksoon and {Song}, Donguk and {Park}, Jongyeob and {Park}, Eunsu and {Lee}, Harim and {Jeong}, Hyun-Jin and {Kang}, Jihye and {Park}, Jinhye and {Yi}, Kangwoo and {Cho}, Il-Hyun and {Na}, Hyeonock},
        title = "{Scientific Perspectives of the Heliophysics L4 Mission by Remote-Sensing Observations}",
      journal = {J.~Korean Astron.~Soc.},
         year = 2024,
        month = mar,
       volume = {57},
        pages = {35-44},
          doi = {10.5303/JKAS.2024.57.1.35},
       adsurl = {https://ui.adsabs.harvard.edu/abs/2024JKAS...57...35M}
}

@ARTICLE{steinberg96,
       author = {{Steinberg}, J.~T. and {Lazarus}, A.~J. and {Ogilvie}, K.~W. and {Lepping}, R. and {Byrnes}, J.},
        title = "{Differential flow between solar wind protons and alpha particles: First WIND observations}",
      journal = {\grl},
         year = 1996,
        month = may,
       volume = {23},
       number = {10},
        pages = {1183-1186},
          doi = {10.1029/96GL00628},
       adsurl = {https://ui.adsabs.harvard.edu/abs/1996GeoRL..23.1183S}
}

@ARTICLE{carlson82,
       author = {{Carlson}, C.~W. and {Curtis}, D.~W. and {Paschmann}, G. and {Michael}, W.},
        title = "{An instrument for rapidly measuring plasma distribution functions with high resolution}",
      journal = {Adv.~Space Res.},
         year = 1982,
        month = jan,
       volume = {2},
       number = {7},
        pages = {67-70},
          doi = {10.1016/0273-1177(82)90151-X},
       adsurl = {https://ui.adsabs.harvard.edu/abs/1982AdSpR...2g..67C}
}

@ARTICLE{gloeckler85,
       author = {{Gloeckler}, G. and {Ipavich}, F.~M. and {Studemann}, W. and {Wilken}, B. and {Hamilton}, D.~C. and {Kremser}, G. and {Hovestadt}, D. and {Gliem}, F. and {Lundgren}, R.~A. and {Rieck}, W. and {Tums}, E.~O. and {Cain}, J.~C. and {Masung}, L.~S. and {Weiss}, W. and {Winterhof}, P.},
        title = "{The Charge-Energy-Mass Spectrometer for 0.3-300 keV/e Ions on the AMPTE CCE}",
      journal = {IEEE Trans.~Geosci.~Remote Sensing},
         year = 1985,
        month = may,
       volume = {23},
       number = {3},
        pages = {234-240},
          doi = {10.1109/TGRS.1985.289519},
       adsurl = {https://ui.adsabs.harvard.edu/abs/1985ITGRS..23..234G}
}

@ARTICLE{moore95,
       author = {{Moore}, T.~E. and {Chappell}, C.~R. and {Chandler}, M.~O. and {Fields}, S.~A. and {Pollock}, C.~J. and {Reasoner}, D.~L. and {Young}, D.~T. and {Burch}, J.~L. and {Eaker}, N. and {Waite}, Jr., J.~H. and {McComas}, D.~J. and {Nordholdt}, J.~E. and {Thomsen}, M.~F. and {Berthelier}, J.~J. and {Robson}, R.},
        title = "{The Thermal Ion Dynamics Experiment and Plasma Source Instrument}",
      journal = {\ssr},
         year = 1995,
        month = feb,
       volume = {71},
       number = {1-4},
        pages = {409-458},
          doi = {10.1007/BF00751337},
       adsurl = {https://ui.adsabs.harvard.edu/abs/1995SSRv...71..409M}
}

@ARTICLE{gershman16,
       author = {{Gershman}, Daniel J. and {Gliese}, Ulrik and {Dorelli}, John C. and {Avanov}, Levon A. and {Barrie}, Alexander C. and {Chornay}, Dennis J. and {MacDonald}, Elizabeth A. and {Holland}, Matthew P. and {Giles}, Barbara L. and {Pollock}, Craig J.},
        title = "{The parameterization of microchannel-plate-based detection systems}",
      journal = {J.~Geophys.~Res.~(Space Phys.)},
         year = 2016,
        month = oct,
       volume = {121},
       number = {10},
        pages = {10,005-10,018},
          doi = {10.1002/2016JA022563},
       adsurl = {https://ui.adsabs.harvard.edu/abs/2016JGRA..12110005G}
}

@ARTICLE{baumgartner76,
       author = {{Baumgartner}, W.~E. and {Huber}, W.~K.},
        title = "{REVIEW ARTICLE: Secondary electron emission multipliers as particle detectors}",
      journal = {J.~Phys.~E Sci.~Instrum.},
         year = 1976,
        month = may,
       volume = {9},
       number = {5},
        pages = {321-330},
          doi = {10.1088/0022-3735/9/5/001},
       adsurl = {https://ui.adsabs.harvard.edu/abs/1976JPhE....9..321B}
}

@ARTICLE{wiza79,
       author = {{Ladislas Wiza}, Joseph},
        title = "{Microchannel plate detectors}",
      journal = {Nucl.~Instrum.~Methods},
         year = 1979,
        month = jun,
       volume = {162},
       number = {1},
        pages = {587-601},
          doi = {10.1016/0029-554X(79)90734-1},
       adsurl = {https://ui.adsabs.harvard.edu/abs/1979NucIM.162..587L}
}

@ARTICLE{funsten15,
       author = {{Funsten}, Herbert O. and {Harper}, Ronnie W. and {Dors}, Eric E. and {Janzen}, Paul A. and {Larsen}, Brian A. and {MacDonald}, Elizabeth A. and {Poston}, David I. and {Ritzau}, Stephen M. and {Skoug}, Ruth M. and {Zurbuchen}, Thomas H.},
        title = "{Comparative Response of Microchannel Plate and Channel Electron Multiplier Detectors to Penetrating Radiation in Space}",
      journal = {IEEE Trans.~Nucl.~Sci.},
         year = 2015,
        month = oct,
       volume = {62},
       number = {5},
        pages = {2283-2293},
          doi = {10.1109/TNS.2015.2464174},
       adsurl = {https://ui.adsabs.harvard.edu/abs/2015ITNS...62.2283F}
}

@ARTICLE{owen20,
       author = {{Owen}, C.~J. and {Bruno}, R. and {Livi}, S. and {Louarn}, P. and {Al Janabi}, K. and {Allegrini}, F. and {Amoros}, C. and {Baruah}, R. and {Barthe}, A. and {Berthomier}, M. and {Bordon}, S. and {Brockley-Blatt}, C. and {Brysbaert}, C. and {Capuano}, G. and {Collier}, M. and {DeMarco}, R. and {Fedorov}, A. and {Ford}, J. and {Fortunato}, V. and {Fratter}, I. and {Galvin}, A.~B. and {Hancock}, B. and {Heirtzler}, D. and {Kataria}, D. and {Kistler}, L. and {Lepri}, S.~T. and {Lewis}, G. and {Loeffler}, C. and {Marty}, W. and {Mathon}, R. and {Mayall}, A. and {Mele}, G. and {Ogasawara}, K. and {Orlandi}, M. and {Pacros}, A. and {Penou}, E. and {Persyn}, S. and {Petiot}, M. and {Phillips}, M. and {P{\v{r}}ech}, L. and {Raines}, J.~M. and {Reden}, M. and {Rouillard}, A.~P. and {Rousseau}, A. and {Rubiella}, J. and {Seran}, H. and {Spencer}, A. and {Thomas}, J.~W. and {Trevino}, J. and {Verscharen}, D. and {Wurz}, P. and {Alapide}, A. and {Amoruso}, L. and {Andr{\'e}}, N. and {Anekallu}, C. and {Arciuli}, V. and {Arnett}, K.~L. and {Ascolese}, R. and {Bancroft}, C. and {Bland}, P. and {Brysch}, M. and {Calvanese}, R. and {Castronuovo}, M. and {{\v{C}}erm{\'a}k}, I. and {Chornay}, D. and {Clemens}, S. and {Coker}, J. and {Collinson}, G. and {D'Amicis}, R. and {Dandouras}, I. and {Darnley}, R. and {Davies}, D. and {Davison}, G. and {De Los Santos}, A. and {Devoto}, P. and {Dirks}, G. and {Edlund}, E. and {Fazakerley}, A. and {Ferris}, M. and {Frost}, C. and {Fruit}, G. and {Garat}, C. and {G{\'e}not}, V. and {Gibson}, W. and {Gilbert}, J.~A. and {de Giosa}, V. and {Gradone}, S. and {Hailey}, M. and {Horbury}, T.~S. and {Hunt}, T. and {Jacquey}, C. and {Johnson}, M. and {Lavraud}, B. and {Lawrenson}, A. and {Leblanc}, F. and {Lockhart}, W. and {Maksimovic}, M. and {Malpus}, A. and {Marcucci}, F. and {Mazelle}, C. and {Monti}, F. and {Myers}, S. and {Nguyen}, T. and {Rodriguez-Pacheco}, J. and {Phillips}, I. and {Popecki}, M. and {Rees}, K. and {Rogacki}, S.~A. and {Ruane}, K. and {Rust}, D. and {Salatti}, M. and {Sauvaud}, J.~A. and {Stakhiv}, M.~O. and {Stange}, J. and {Stubbs}, T. and {Taylor}, T. and {Techer}, J. -D. and {Terrier}, G. and {Thibodeaux}, R. and {Urdiales}, C. and {Varsani}, A. and {Walsh}, A.~P. and {Watson}, G. and {Wheeler}, P. and {Willis}, G. and {Wimmer-Schweingruber}, R.~F. and {Winter}, B. and {Yardley}, J. and {Zouganelis}, I.},
        title = "{The Solar Orbiter Solar Wind Analyser (SWA) suite}",
      journal = {\aap},
         year = 2020,
        month = oct,
       volume = {642},
          eid = {A16},
        pages = {A16},
          doi = {10.1051/0004-6361/201937259},
       adsurl = {https://ui.adsabs.harvard.edu/abs/2020A&A...642A..16O}
}

@ARTICLE{kasper16,
       author = {{Kasper}, Justin C. and {Abiad}, Robert and {Austin}, Gerry and {Balat-Pichelin}, Marianne and {Bale}, Stuart D. and {Belcher}, John W. and {Berg}, Peter and {Bergner}, Henry and {Berthomier}, Matthieu and {Bookbinder}, Jay and {Brodu}, Etienne and {Caldwell}, David and {Case}, Anthony W. and {Chandran}, Benjamin D.~G. and {Cheimets}, Peter and {Cirtain}, Jonathan W. and {Cranmer}, Steven R. and {Curtis}, David W. and {Daigneau}, Peter and {Dalton}, Greg and {Dasgupta}, Brahmananda and {DeTomaso}, David and {Diaz-Aguado}, Millan and {Djordjevic}, Blagoje and {Donaskowski}, Bill and {Effinger}, Michael and {Florinski}, Vladimir and {Fox}, Nichola and {Freeman}, Mark and {Gallagher}, Dennis and {Gary}, S. Peter and {Gauron}, Tom and {Gates}, Richard and {Goldstein}, Melvin and {Golub}, Leon and {Gordon}, Dorothy A. and {Gurnee}, Reid and {Guth}, Giora and {Halekas}, Jasper and {Hatch}, Ken and {Heerikuisen}, Jacob and {Ho}, George and {Hu}, Qiang and {Johnson}, Greg and {Jordan}, Steven P. and {Korreck}, Kelly E. and {Larson}, Davin and {Lazarus}, Alan J. and {Li}, Gang and {Livi}, Roberto and {Ludlam}, Michael and {Maksimovic}, Milan and {McFadden}, James P. and {Marchant}, William and {Maruca}, Bennet A. and {McComas}, David J. and {Messina}, Luciana and {Mercer}, Tony and {Park}, Sang and {Peddie}, Andrew M. and {Pogorelov}, Nikolai and {Reinhart}, Matthew J. and {Richardson}, John D. and {Robinson}, Miles and {Rosen}, Irene and {Skoug}, Ruth M. and {Slagle}, Amanda and {Steinberg}, John T. and {Stevens}, Michael L. and {Szabo}, Adam and {Taylor}, Ellen R. and {Tiu}, Chris and {Turin}, Paul and {Velli}, Marco and {Webb}, Gary and {Whittlesey}, Phyllis and {Wright}, Ken and {Wu}, S.~T. and {Zank}, Gary},
        title = "{Solar Wind Electrons Alphas and Protons (SWEAP) Investigation: Design of the Solar Wind and Coronal Plasma Instrument Suite for Solar Probe Plus}",
      journal = {\ssr},
         year = 2016,
        month = dec,
       volume = {204},
       number = {1-4},
        pages = {131-186},
          doi = {10.1007/s11214-015-0206-3},
       adsurl = {https://ui.adsabs.harvard.edu/abs/2016SSRv..204..131K}
}

@ARTICLE{collinson10,
       author = {{Collinson}, Glyn A. and {Kataria}, Dhiren O.},
        title = "{On variable geometric factor systems for top-hat electrostatic space plasma analyzers}",
      journal = {Measurement Science and Technology},
         year = 2010,
        month = oct,
       volume = {21},
       number = {10},
          eid = {105903},
        pages = {105903},
          doi = {10.1088/0957-0233/21/10/105903},
       adsurl = {https://ui.adsabs.harvard.edu/abs/2010MeScT..21j5903C}
}

@ARTICLE{cara17,
       author = {{Cara}, Antoine and {Lavraud}, Benoit and {Fedorov}, Andrei and {De Keyser}, Johan and {DeMarco}, Rossana and {Marcucci}, M. Federica and {Valentini}, Francesco and {Servidio}, Sergio and {Bruno}, Roberto},
        title = "{Electrostatic analyzer design for solar wind proton measurements with high temporal, energy, and angular resolutions}",
      journal = {J.~Geophys.~Res.~(Space Phys.)},
         year = 2017,
        month = feb,
       volume = {122},
       number = {2},
        pages = {1439-1450},
          doi = {10.1002/2016JA023269},
       adsurl = {https://ui.adsabs.harvard.edu/abs/2017JGRA..122.1439C}
}

@ARTICLE{kessel89,
       author = {{Kessel}, R.~L. and {Johnstone}, A.~D. and {Coates}, A.~J. and {Gowen}, R.~A.},
        title = "{Space plasma measurements with ion instruments}",
      journal = {Rev.~Sci.~Instrum.},
         year = 1989,
        month = dec,
       volume = {60},
        pages = {3750-3761},
          doi = {10.1063/1.1141075},
       adsurl = {https://ui.adsabs.harvard.edu/abs/1989RScI...60.3750K}
}

@ARTICLE{nicolaou20,
       author = {{Nicolaou}, G. and {Wicks}, R.~T. and {Rae}, I.~J. and {Kataria}, D.~O.},
        title = "{Evaluating the Performance of a Plasma Analyzer for a Space Weather Monitor Mission Concept}",
      journal = {Space Weather},
         year = 2020,
        month = dec,
       volume = {18},
       number = {12},
          eid = {e2020SW002559},
        pages = {e2020SW002559},
          doi = {10.1029/2020SW002559},
       adsurl = {https://ui.adsabs.harvard.edu/abs/2020SpWea..1802559N}
}

@ARTICLE{nicolaou25,
       author = {{Nicolaou}, G. and {Ioannou}, C. and {Owen}, C.~J. and {Verscharen}, D. and {Fedorov}, A. and {Louarn}, P.},
        title = "{How does the limited resolution of space plasma analyzers affect the accuracy of space plasma measurements?}",
      journal = {Rev.~Sci.~Instrum.},
         year = 2025,
        month = jul,
       volume = {96},
       number = {7},
          eid = {075203},
        pages = {075203},
          doi = {10.1063/5.0218667},
archivePrefix = {arXiv},
       eprint = {2505.09869},
 primaryClass = {physics.space-ph},
       adsurl = {https://ui.adsabs.harvard.edu/abs/2025RScI...96g5203N}
}

@ARTICLE{nicolaou24,
       author = {{Nicolaou}, G. and {Livadiotis}, G. and {Sarlis}, N. and {Ioannou}, C.},
        title = "{Resolving velocity distribution function parameters from observations with significant Poisson statistical uncertainty}",
      journal = {RAS Techniques and Instruments},
         year = 2024,
        month = jan,
       volume = {3},
       number = {1},
        pages = {874-878},
          doi = {10.1093/rasti/rzae059},
       adsurl = {https://ui.adsabs.harvard.edu/abs/2024RASTI...3..874N}
}

@ARTICLE{nicolaou23,
       author = {{Nicolaou}, Georgios},
        title = "{Effects of noise on the accuracy of plasma bulk parameters derived from velocity moments of in-situ observations}",
      journal = {\apss},
         year = 2023,
        month = jan,
       volume = {368},
       number = {1},
          eid = {3},
        pages = {3},
          doi = {10.1007/s10509-022-04157-z},
       adsurl = {https://ui.adsabs.harvard.edu/abs/2023Ap&SS.368....3N}
}

@ARTICLE{nicolaou22,
       author = {{Nicolaou}, Georgios and {Haythornthwaite}, Richard P. and {Coates}, Andrew J.},
        title = "{Resolving Space Plasma Species With Electrostatic Analyzers}",
      journal = {Front.~Astron.~Space Sci.},
         year = 2022,
        month = jun,
       volume = {9},
          eid = {861433},
        pages = {861433},
          doi = {10.3389/fspas.2022.861433},
       adsurl = {https://ui.adsabs.harvard.edu/abs/2022FrASS...9.1433N}
}

@ARTICLE{lavraud16,
       author = {{Lavraud}, Benoit and {Larson}, Davin E.},
        title = "{Correcting moments of in situ particle distribution functions for spacecraft electrostatic charging}",
      journal = {J.~Geophys.~Res.~(Space Phys.)},
         year = 2016,
        month = sep,
       volume = {121},
       number = {9},
        pages = {8462-8474},
          doi = {10.1002/2016JA022591},
       adsurl = {https://ui.adsabs.harvard.edu/abs/2016JGRA..121.8462L}
}

@ARTICLE{wilson15,
       author = {{Wilson}, R.~J.},
        title = "{Error analysis for numerical estimates of space plasma parameters}",
      journal = {Earth Space Sci.},
         year = 2015,
        month = jun,
       volume = {2},
       number = {6},
        pages = {201-222},
          doi = {10.1002/2014EA000090},
       adsurl = {https://ui.adsabs.harvard.edu/abs/2015E&SS....2..201W}
}

@ARTICLE{wilson22,
       author = {{Wilson}, III, Lynn B. and {Goodrich}, Katherine A. and {Turner}, Drew L. and {Cohen}, Ian J. and {Whittlesey}, Phyllis L. and {Schwartz}, Steven J.},
        title = "{The need for accurate measurements of thermal velocity distribution functions in the solar wind}",
      journal = {Front.~Astron.~Space Sci.},
         year = 2022,
        month = nov,
       volume = {9},
          eid = {369},
        pages = {369},
          doi = {10.3389/fspas.2022.1063841},
       adsurl = {https://ui.adsabs.harvard.edu/abs/2022FrASS...963841W}
}

@ARTICLE{parent24,
       author = {{Parent}, P. -Y. and {Verscharen}, D. and {Nicolaou}, G. and {Owen}, C.~J.},
        title = "{Microchannel plate response to solar energetic particles and consequences for solar-wind measurements on ESA's Vigil mission}",
      journal = {RAS Techniques and Instruments},
         year = 2024,
        month = jan,
       volume = {3},
       number = {1},
        pages = {844-852},
          doi = {10.1093/rasti/rzae053},
       adsurl = {https://ui.adsabs.harvard.edu/abs/2024RASTI...3..844P}
}

@ARTICLE{mueller20,
       author = {{M{\"u}ller}, D. and {St. Cyr}, O.~C. and {Zouganelis}, I. and {Gilbert}, H.~R. and {Marsden}, R. and {Nieves-Chinchilla}, T. and {Antonucci}, E. and {Auch{\`e}re}, F. and {Berghmans}, D. and {Horbury}, T.~S. and {Howard}, R.~A. and {Krucker}, S. and {Maksimovic}, M. and {Owen}, C.~J. and {Rochus}, P. and {Rodriguez-Pacheco}, J. and {Romoli}, M. and {Solanki}, S.~K. and {Bruno}, R. and {Carlsson}, M. and {Fludra}, A. and {Harra}, L. and {Hassler}, D.~M. and {Livi}, S. and {Louarn}, P. and {Peter}, H. and {Sch{\"u}hle}, U. and {Teriaca}, L. and {del Toro Iniesta}, J.~C. and {Wimmer-Schweingruber}, R.~F. and {Marsch}, E. and {Velli}, M. and {De Groof}, A. and {Walsh}, A. and {Williams}, D.},
        title = "{The Solar Orbiter mission. Science overview}",
      journal = {\aap},
         year = 2020,
        month = oct,
       volume = {642},
          eid = {A1},
        pages = {A1},
          doi = {10.1051/0004-6361/202038467},
archivePrefix = {arXiv},
       eprint = {2009.00861},
 primaryClass = {astro-ph.SR},
       adsurl = {https://ui.adsabs.harvard.edu/abs/2020A&A...642A...1M}
}

@ARTICLE{garcia21,
       author = {{Garc{\'\i}a Marirrodriga}, C. and {Pacros}, A. and {Strandmoe}, S. and {Arcioni}, M. and {Arts}, A. and {Ashcroft}, C. and {Ayache}, L. and {Bonnefous}, Y. and {Brahimi}, N. and {Cipriani}, F. and {Damasio}, C. and {De Jong}, P. and {D{\'e}prez}, G. and {Fahmy}, S. and {Fels}, R. and {Fiebrich}, J. and {Hass}, C. and {Hern{\'a}ndez}, C. and {Icardi}, L. and {Junge}, A. and {Kletzkine}, P. and {Laget}, P. and {Le Deuff}, Y. and {Liebold}, F. and {Lodiot}, S. and {Marliani}, F. and {Mascarello}, M. and {M{\"u}ller}, D. and {Oganessian}, A. and {Olivier}, P. and {Palombo}, E. and {Philippe}, C. and {Ragnit}, U. and {Ramachandran}, J. and {S{\'a}nchez P{\'e}rez}, J.~M. and {Stienstra}, M.~M. and {Th{\"u}rey}, S. and {Urwin}, A. and {Wirth}, K. and {Zouganelis}, I.},
        title = "{Solar Orbiter: Mission and spacecraft design}",
      journal = {\aap},
         year = 2021,
        month = feb,
       volume = {646},
          eid = {A121},
        pages = {A121},
          doi = {10.1051/0004-6361/202038519},
       adsurl = {https://ui.adsabs.harvard.edu/abs/2021A&A...646A.121G}
}

@ARTICLE{owen21,
       author = {{Owen}, C.~J. and {Kataria}, D.~O. and {Ber{\v{c}}i{\v{c}}}, L. and {Horbury}, T.~S. and {Berthomier}, M. and {Verscharen}, D. and {Bruno}, R. and {Livi}, S. and {Louarn}, P. and {Anekallu}, C. and {Kelly}, C.~W. and {Lewis}, G.~R. and {Watson}, G. and {Fortunato}, V. and {Mele}, G. and {Nicolaou}, G. and {Wicks}, R.~T. and {O'Brien}, H. and {Evans}, V. and {Angelini}, V.},
        title = "{High-cadence measurements of electron pitch-angle distributions from Solar Orbiter SWA-EAS burst mode operations}",
      journal = {\aap},
         year = 2021,
        month = dec,
       volume = {656},
          eid = {L9},
        pages = {L9},
          doi = {10.1051/0004-6361/202140959},
       adsurl = {https://ui.adsabs.harvard.edu/abs/2021A&A...656L...9O}
}

@ARTICLE{bercic21,
       author = {{Ber{\v{c}}i{\v{c}}}, L. and {Verscharen}, D. and {Owen}, C.~J. and {Colomban}, L. and {Kretzschmar}, M. and {Chust}, T. and {Maksimovic}, M. and {Kataria}, D.~O. and {Anekallu}, C. and {Behar}, E. and {Berthomier}, M. and {Bruno}, R. and {Fortunato}, V. and {Kelly}, C.~W. and {Khotyaintsev}, Y.~V. and {Lewis}, G.~R. and {Livi}, S. and {Louarn}, P. and {Mele}, G. and {Nicolaou}, G. and {Watson}, G. and {Wicks}, R.~T.},
        title = "{Whistler instability driven by the sunward electron deficit in the solar wind. High-cadence Solar Orbiter observations}",
      journal = {\aap},
         year = 2021,
        month = dec,
       volume = {656},
          eid = {A31},
        pages = {A31},
          doi = {10.1051/0004-6361/202140970},
archivePrefix = {arXiv},
       eprint = {2107.10645},
 primaryClass = {physics.space-ph},
       adsurl = {https://ui.adsabs.harvard.edu/abs/2021A&A...656A..31B}
}

@ARTICLE{feldman75,
       author = {{Feldman}, W.~C. and {Asbridge}, J.~R. and {Bame}, S.~J. and {Montgomery}, M.~D. and {Gary}, S.~P.},
        title = "{Solar wind electrons}",
      journal = {\jgr},
         year = 1975,
        month = nov,
       volume = {80},
       number = {31},
        pages = {4181},
          doi = {10.1029/JA080i031p04181},
       adsurl = {https://ui.adsabs.harvard.edu/abs/1975JGR....80.4181F}
}

@ARTICLE{pilipp87,
       author = {{Pilipp}, W.~G. and {Miggenrieder}, H. and {Montgomery}, M.~D. and {M{\"u}hlh{\"a}user}, K. -H. and {Rosenbauer}, H. and {Schwenn}, R.},
        title = "{Characteristics of electron velocity distribution functions in the solar wind derived from the helios plasma experiment}",
      journal = {\jgr},
         year = 1987,
        month = feb,
       volume = {92},
       number = {A2},
        pages = {1075-1092},
          doi = {10.1029/JA092iA02p01075},
       adsurl = {https://ui.adsabs.harvard.edu/abs/1987JGR....92.1075P}
}

@ARTICLE{lin98,
       author = {{Lin}, R.~P.},
        title = "{WIND Observations of Suprathermal Electrons in the Interplanetary Medium}",
      journal = {\ssr},
         year = 1998,
        month = jul,
       volume = {86},
        pages = {61-78},
          doi = {10.1023/A:1005048428480},
       adsurl = {https://ui.adsabs.harvard.edu/abs/1998SSRv...86...61L}
}

@ARTICLE{salem23,
       author = {{Salem}, Chadi S. and {Pulupa}, Marc and {Bale}, Stuart D. and {Verscharen}, Daniel},
        title = "{Precision electron measurements in the solar wind at 1 au from NASA's Wind spacecraft}",
      journal = {\aap},
         year = 2023,
        month = jul,
       volume = {675},
          eid = {A162},
        pages = {A162},
          doi = {10.1051/0004-6361/202141816},
archivePrefix = {arXiv},
       eprint = {2107.08125},
 primaryClass = {physics.space-ph},
       adsurl = {https://ui.adsabs.harvard.edu/abs/2023A&A...675A.162S}
}

@ARTICLE{halekas21,
       author = {{Halekas}, J.~S. and {Ber{\v{c}}i{\v{c}}}, L. and {Whittlesey}, P. and {Larson}, D.~E. and {Livi}, R. and {Berthomier}, M. and {Kasper}, J.~C. and {Case}, A.~W. and {Stevens}, M.~L. and {Bale}, S.~D. and {MacDowall}, R.~J. and {Pulupa}, M.~P.},
        title = "{The Sunward Electron Deficit: A Telltale Sign of the Sun's Electric Potential}",
      journal = {\apj},
         year = 2021,
        month = jul,
       volume = {916},
       number = {1},
          eid = {16},
        pages = {16},
          doi = {10.3847/1538-4357/ac096e},
       adsurl = {https://ui.adsabs.harvard.edu/abs/2021ApJ...916...16H}
}

@ARTICLE{lemaire71,
       author = {{Lemaire}, J. and {Scherer}, M.},
        title = "{Kinetic models of the solar wind}",
      journal = {\jgr},
         year = 1971,
        month = jan,
       volume = {76},
       number = {31},
        pages = {7479},
          doi = {10.1029/JA076i031p07479},
       adsurl = {https://ui.adsabs.harvard.edu/abs/1971JGR....76.7479L}
}

@ARTICLE{maksimovic01,
       author = {{Maksimovic}, Milan and {Pierrard}, Viviane and {Lemaire}, Joseph},
        title = "{On the Exospheric Approach for the Solar Wind Acceleration}",
      journal = {\apss},
         year = 2001,
        month = jun,
       volume = {277},
        pages = {181-187},
          doi = {10.1023/A:1012250027289},
       adsurl = {https://ui.adsabs.harvard.edu/abs/2001Ap&SS.277..181M}
}

@ARTICLE{bercic21b,
       author = {{Ber{\v{c}}i{\v{c}}}, Laura and {Maksimovi{\'c}}, Milan and {Halekas}, Jasper S. and {Landi}, Simone and {Owen}, Christopher J. and {Verscharen}, Daniel and {Larson}, Davin and {Whittlesey}, Phyllis and {Badman}, Samuel T. and {Bale}, Stuart. D. and {Case}, Anthony W. and {Goetz}, Keith and {Harvey}, Peter R. and {Kasper}, Justin C. and {Korreck}, Kelly E. and {Livi}, Roberto and {MacDowall}, Robert J. and {Malaspina}, David M. and {Pulupa}, Marc and {Stevens}, Michael L.},
        title = "{Ambipolar Electric Field and Potential in the Solar Wind Estimated from Electron Velocity Distribution Functions}",
      journal = {\apj},
         year = 2021,
        month = nov,
       volume = {921},
       number = {1},
          eid = {83},
        pages = {83},
          doi = {10.3847/1538-4357/ac1f1c},
archivePrefix = {arXiv},
       eprint = {2108.08528},
 primaryClass = {astro-ph.SR},
       adsurl = {https://ui.adsabs.harvard.edu/abs/2021ApJ...921...83B}
}

@ARTICLE{micera25,
       author = {{Micera}, Alfredo and {Verscharen}, Daniel and {Coburn}, Jesse T. and {Innocenti}, Maria Elena},
        title = "{Quasi-parallel Antisunward-propagating Whistler Waves Associated with the Electron Deficit in the Near-Sun Solar Wind: Particle-in-cell Simulation}",
      journal = {\apj},
         year = 2025,
        month = feb,
       volume = {979},
       number = {2},
          eid = {226},
        pages = {226},
          doi = {10.3847/1538-4357/ada3d7},
archivePrefix = {arXiv},
       eprint = {2501.01331},
 primaryClass = {astro-ph.SR},
       adsurl = {https://ui.adsabs.harvard.edu/abs/2025ApJ...979..226M}
}

@ARTICLE{fox16,
       author = {{Fox}, N.~J. and {Velli}, M.~C. and {Bale}, S.~D. and {Decker}, R. and {Driesman}, A. and {Howard}, R.~A. and {Kasper}, J.~C. and {Kinnison}, J. and {Kusterer}, M. and {Lario}, D. and {Lockwood}, M.~K. and {McComas}, D.~J. and {Raouafi}, N.~E. and {Szabo}, A.},
        title = "{The Solar Probe Plus Mission: Humanity's First Visit to Our Star}",
      journal = {\ssr},
         year = 2016,
        month = dec,
       volume = {204},
       number = {1-4},
        pages = {7-48},
          doi = {10.1007/s11214-015-0211-6},
       adsurl = {https://ui.adsabs.harvard.edu/abs/2016SSRv..204....7F}
}

@ARTICLE{livi22,
       author = {{Livi}, Roberto and {Larson}, Davin E. and {Kasper}, Justin C. and {Abiad}, Robert and {Case}, A.~W. and {Klein}, Kristopher G. and {Curtis}, David W. and {Dalton}, Gregory and {Stevens}, Michael and {Korreck}, Kelly E. and {Ho}, George and {Robinson}, Miles and {Tiu}, Chris and {Whittlesey}, Phyllis L. and {Verniero}, Jaye L. and {Halekas}, Jasper and {McFadden}, James and {Marckwordt}, Mario and {Slagle}, Amanda and {Abatcha}, Mamuda and {Rahmati}, Ali and {McManus}, Michael D.},
        title = "{The Solar Probe ANalyzer-Ions on the Parker Solar Probe}",
      journal = {\apj},
         year = 2022,
        month = oct,
       volume = {938},
       number = {2},
          eid = {138},
        pages = {138},
          doi = {10.3847/1538-4357/ac93f5},
       adsurl = {https://ui.adsabs.harvard.edu/abs/2022ApJ...938..138L}
}

@ARTICLE{whittlesey20,
       author = {{Whittlesey}, Phyllis L. and {Larson}, Davin E. and {Kasper}, Justin C. and {Halekas}, Jasper and {Abatcha}, Mamuda and {Abiad}, Robert and {Berthomier}, M. and {Case}, A.~W. and {Chen}, Jianxin and {Curtis}, David W. and {Dalton}, Gregory and {Klein}, Kristopher G. and {Korreck}, Kelly E. and {Livi}, Roberto and {Ludlam}, Michael and {Marckwordt}, Mario and {Rahmati}, Ali and {Robinson}, Miles and {Slagle}, Amanda and {Stevens}, M.~L. and {Tiu}, Chris and {Verniero}, J.~L.},
        title = "{The Solar Probe ANalyzers{\textemdash}Electrons on the Parker Solar Probe}",
      journal = {\apjs},
         year = 2020,
        month = feb,
       volume = {246},
       number = {2},
          eid = {74},
        pages = {74},
          doi = {10.3847/1538-4365/ab7370},
archivePrefix = {arXiv},
       eprint = {2002.04080},
 primaryClass = {astro-ph.IM},
       adsurl = {https://ui.adsabs.harvard.edu/abs/2020ApJS..246...74W}
}

@ARTICLE{case20,
       author = {{Case}, A.~W. and {Kasper}, Justin C. and {Stevens}, Michael L. and {Korreck}, Kelly E. and {Paulson}, Kristoff and {Daigneau}, Peter and {Caldwell}, Dave and {Freeman}, Mark and {Henry}, Thayne and {Klingensmith}, Brianna and {Bookbinder}, J.~A. and {Robinson}, Miles and {Berg}, Peter and {Tiu}, Chris and {Wright}, Jr., K.~H. and {Reinhart}, Matthew J. and {Curtis}, David and {Ludlam}, Michael and {Larson}, Davin and {Whittlesey}, Phyllis and {Livi}, Roberto and {Klein}, Kristopher G. and {Martinovi{\'c}}, Mihailo M.},
        title = "{The Solar Probe Cup on the Parker Solar Probe}",
      journal = {\apjs},
         year = 2020,
        month = feb,
       volume = {246},
       number = {2},
          eid = {43},
        pages = {43},
          doi = {10.3847/1538-4365/ab5a7b},
archivePrefix = {arXiv},
       eprint = {1912.02581},
 primaryClass = {astro-ph.IM},
       adsurl = {https://ui.adsabs.harvard.edu/abs/2020ApJS..246...43C}
}

@ARTICLE{verniero22,
       author = {{Verniero}, J.~L. and {Chandran}, B.~D.~G. and {Larson}, D.~E. and {Paulson}, K. and {Alterman}, B.~L. and {Badman}, S. and {Bale}, S.~D. and {Bonnell}, J.~W. and {Bowen}, T.~A. and {de Wit}, T. Dudok and {Kasper}, J.~C. and {Klein}, K.~G. and {Lichko}, E. and {Livi}, R. and {McManus}, M.~D. and {Rahmati}, A. and {Verscharen}, D. and {Walters}, J. and {Whittlesey}, P.~L.},
        title = "{Strong Perpendicular Velocity-space Diffusion in Proton Beams Observed by Parker Solar Probe}",
      journal = {\apj},
         year = 2022,
        month = jan,
       volume = {924},
       number = {2},
          eid = {112},
        pages = {112},
          doi = {10.3847/1538-4357/ac36d5},
archivePrefix = {arXiv},
       eprint = {2110.08912},
 primaryClass = {astro-ph.SR},
       adsurl = {https://ui.adsabs.harvard.edu/abs/2022ApJ...924..112V}
}

@ARTICLE{bale16,
       author = {{Bale}, S.~D. and {Goetz}, K. and {Harvey}, P.~R. and {Turin}, P. and {Bonnell}, J.~W. and {Dudok de Wit}, T. and {Ergun}, R.~E. and {MacDowall}, R.~J. and {Pulupa}, M. and {Andre}, M. and {Bolton}, M. and {Bougeret}, J. -L. and {Bowen}, T.~A. and {Burgess}, D. and {Cattell}, C.~A. and {Chandran}, B.~D.~G. and {Chaston}, C.~C. and {Chen}, C.~H.~K. and {Choi}, M.~K. and {Connerney}, J.~E. and {Cranmer}, S. and {Diaz-Aguado}, M. and {Donakowski}, W. and {Drake}, J.~F. and {Farrell}, W.~M. and {Fergeau}, P. and {Fermin}, J. and {Fischer}, J. and {Fox}, N. and {Glaser}, D. and {Goldstein}, M. and {Gordon}, D. and {Hanson}, E. and {Harris}, S.~E. and {Hayes}, L.~M. and {Hinze}, J.~J. and {Hollweg}, J.~V. and {Horbury}, T.~S. and {Howard}, R.~A. and {Hoxie}, V. and {Jannet}, G. and {Karlsson}, M. and {Kasper}, J.~C. and {Kellogg}, P.~J. and {Kien}, M. and {Klimchuk}, J.~A. and {Krasnoselskikh}, V.~V. and {Krucker}, S. and {Lynch}, J.~J. and {Maksimovic}, M. and {Malaspina}, D.~M. and {Marker}, S. and {Martin}, P. and {Martinez-Oliveros}, J. and {McCauley}, J. and {McComas}, D.~J. and {McDonald}, T. and {Meyer-Vernet}, N. and {Moncuquet}, M. and {Monson}, S.~J. and {Mozer}, F.~S. and {Murphy}, S.~D. and {Odom}, J. and {Oliverson}, R. and {Olson}, J. and {Parker}, E.~N. and {Pankow}, D. and {Phan}, T. and {Quataert}, E. and {Quinn}, T. and {Ruplin}, S.~W. and {Salem}, C. and {Seitz}, D. and {Sheppard}, D.~A. and {Siy}, A. and {Stevens}, K. and {Summers}, D. and {Szabo}, A. and {Timofeeva}, M. and {Vaivads}, A. and {Velli}, M. and {Yehle}, A. and {Werthimer}, D. and {Wygant}, J.~R.},
        title = "{The FIELDS Instrument Suite for Solar Probe Plus. Measuring the Coronal Plasma and Magnetic Field, Plasma Waves and Turbulence, and Radio Signatures of Solar Transients}",
      journal = {\ssr},
         year = 2016,
        month = dec,
       volume = {204},
       number = {1-4},
        pages = {49-82},
          doi = {10.1007/s11214-016-0244-5},
       adsurl = {https://ui.adsabs.harvard.edu/abs/2016SSRv..204...49B}
}

@ARTICLE{maksimovic20,
       author = {{Maksimovic}, M. and {Bale}, S.~D. and {Chust}, T. and {Khotyaintsev}, Y. and {Krasnoselskikh}, V. and {Kretzschmar}, M. and {Plettemeier}, D. and {Rucker}, H.~O. and {Sou{\v{c}}ek}, J. and {Steller}, M. and {{\v{S}}tver{\'a}k}, {\v{S}}. and {Tr{\'a}vn{\'\i}{\v{c}}ek}, P. and {Vaivads}, A. and {Chaintreuil}, S. and {Dekkali}, M. and {Alexandrova}, O. and {Astier}, P. -A. and {Barbary}, G. and {B{\'e}rard}, D. and {Bonnin}, X. and {Boughedada}, K. and {Cecconi}, B. and {Chapron}, F. and {Chariet}, M. and {Collin}, C. and {de Conchy}, Y. and {Dias}, D. and {Gu{\'e}guen}, L. and {Lamy}, L. and {Leray}, V. and {Lion}, S. and {Malac-Allain}, L.~R. and {Matteini}, L. and {Nguyen}, Q.~N. and {Pantellini}, F. and {Parisot}, J. and {Plasson}, P. and {Thijs}, S. and {Vecchio}, A. and {Fratter}, I. and {Bellouard}, E. and {Lorf{\`e}vre}, E. and {Danto}, P. and {Julien}, S. and {Guilhem}, E. and {Fiachetti}, C. and {Sanisidro}, J. and {Laffaye}, C. and {Gonzalez}, F. and {Pontet}, B. and {Qu{\'e}ruel}, N. and {Jannet}, G. and {Fergeau}, P. and {Brochot}, J. -Y. and {Cassam-Chenai}, G. and {Dudok de Wit}, T. and {Timofeeva}, M. and {Vincent}, T. and {Agrapart}, C. and {Delory}, G.~T. and {Turin}, P. and {Jeandet}, A. and {Leroy}, P. and {Pellion}, J. -C. and {Bouzid}, V. and {Katra}, B. and {Piberne}, R. and {Recart}, W. and {Santol{\'\i}k}, O. and {Kolma{\v{s}}ov{\'a}}, I. and {Krupa{\v{r}}}, V. and {Krupa{\v{r}}ov{\'a}}, O. and {P{\'\i}{\v{s}}a}, D. and {Uhl{\'\i}{\v{r}}}, L. and {L{\'a}n}, R. and {Ba{\v{s}}e}, J. and {Ahl{\`e}n}, L. and {Andr{\'e}}, M. and {Bylander}, L. and {Cripps}, V. and {Cully}, C. and {Eriksson}, A. and {Jansson}, S. -E. and {Johansson}, E.~P.~G. and {Karlsson}, T. and {Puccio}, W. and {B{\v{r}}{\'\i}nek}, J. and {{\"O}ttacher}, H. and {Panchenko}, M. and {Berthomier}, M. and {Goetz}, K. and {Hellinger}, P. and {Horbury}, T.~S. and {Issautier}, K. and {Kontar}, E. and {Krucker}, S. and {Le Contel}, O. and {Louarn}, P. and {Martinovi{\'c}}, M. and {Owen}, C.~J. and {Retino}, A. and {Rodr{\'\i}guez-Pacheco}, J. and {Sahraoui}, F. and {Wimmer-Schweingruber}, R.~F. and {Zaslavsky}, A. and {Zouganelis}, I.},
        title = "{The Solar Orbiter Radio and Plasma Waves (RPW) instrument}",
      journal = {\aap},
         year = 2020,
        month = oct,
       volume = {642},
          eid = {A12},
        pages = {A12},
          doi = {10.1051/0004-6361/201936214},
       adsurl = {https://ui.adsabs.harvard.edu/abs/2020A&A...642A..12M}
}

@ARTICLE{horbury20,
       author = {{Horbury}, T.~S. and {O'Brien}, H. and {Carrasco Blazquez}, I. and {Bendyk}, M. and {Brown}, P. and {Hudson}, R. and {Evans}, V. and {Oddy}, T.~M. and {Carr}, C.~M. and {Beek}, T.~J. and {Cupido}, E. and {Bhattacharya}, S. and {Dominguez}, J. -A. and {Matthews}, L. and {Myklebust}, V.~R. and {Whiteside}, B. and {Bale}, S.~D. and {Baumjohann}, W. and {Burgess}, D. and {Carbone}, V. and {Cargill}, P. and {Eastwood}, J. and {Erd{\"o}s}, G. and {Fletcher}, L. and {Forsyth}, R. and {Giacalone}, J. and {Glassmeier}, K. -H. and {Goldstein}, M.~L. and {Hoeksema}, T. and {Lockwood}, M. and {Magnes}, W. and {Maksimovic}, M. and {Marsch}, E. and {Matthaeus}, W.~H. and {Murphy}, N. and {Nakariakov}, V.~M. and {Owen}, C.~J. and {Owens}, M. and {Rodriguez-Pacheco}, J. and {Richter}, I. and {Riley}, P. and {Russell}, C.~T. and {Schwartz}, S. and {Vainio}, R. and {Velli}, M. and {Vennerstrom}, S. and {Walsh}, R. and {Wimmer-Schweingruber}, R.~F. and {Zank}, G. and {M{\"u}ller}, D. and {Zouganelis}, I. and {Walsh}, A.~P.},
        title = "{The Solar Orbiter magnetometer}",
      journal = {\aap},
         year = 2020,
        month = oct,
       volume = {642},
          eid = {A9},
        pages = {A9},
          doi = {10.1051/0004-6361/201937257},
       adsurl = {https://ui.adsabs.harvard.edu/abs/2020A&A...642A...9H}
}

@ARTICLE{marsch82,
       author = {{Marsch}, E. and {Schwenn}, R. and {Rosenbauer}, H. and {Muehlhaeuser}, K. -H. and {Pilipp}, W. and {Neubauer}, F.~M.},
        title = "{Solar wind protons: Three-dimensional velocity distributions and derived plasma parameters measured between 0.3 and 1 AU}",
      journal = {\jgr},
         year = 1982,
        month = jan,
       volume = {87},
       number = {A1},
        pages = {52-72},
          doi = {10.1029/JA087iA01p00052},
       adsurl = {https://ui.adsabs.harvard.edu/abs/1982JGR....87...52M}
}

@ARTICLE{bruno24,
       author = {{Bruno}, Roberto and {De Marco}, Rossana and {D'Amicis}, Raffaella and {Perrone}, Denise and {Marcucci}, Maria Federica and {Telloni}, Daniele and {Marino}, Raffaele and {Sorriso-Valvo}, Luca and {Fortunato}, Vito and {Mele}, Gennaro and {Monti}, Francesco and {Fedorov}, Andrei and {Louarn}, Philippe and {Owen}, Chris J. and {Livi}, Stefano},
        title = "{Comparative Study of the Kinetic Properties of Proton and Alpha Beams in the Alfv{\'e}nic Wind Observed by SWA-PAS On Board Solar Orbiter}",
      journal = {\apj},
         year = 2024,
        month = jul,
       volume = {969},
       number = {2},
          eid = {106},
        pages = {106},
          doi = {10.3847/1538-4357/ad47b3},
archivePrefix = {arXiv},
       eprint = {2403.10489},
 primaryClass = {astro-ph.SR},
       adsurl = {https://ui.adsabs.harvard.edu/abs/2024ApJ...969..106B}
}

@ARTICLE{durovcova21,
       author = {{{\v{D}}urovcov{\'a}}, Tereza and {{\v{S}}afr{\'a}nkov{\'a}}, Jana and {N{\v{e}}me{\v{c}}ek}, Zden{\v{e}}k},
        title = "{Proton Beam Abundance Variations and Their Relation to Alpha Particle Properties}",
      journal = {\apj},
         year = 2021,
        month = dec,
       volume = {923},
       number = {2},
          eid = {170},
        pages = {170},
          doi = {10.3847/1538-4357/ac2c03},
       adsurl = {https://ui.adsabs.harvard.edu/abs/2021ApJ...923..170D}
}

@ARTICLE{verniero20,
       author = {{Verniero}, J.~L. and {Larson}, D.~E. and {Livi}, R. and {Rahmati}, A. and {McManus}, M.~D. and {Pyakurel}, P. Sharma and {Klein}, K.~G. and {Bowen}, T.~A. and {Bonnell}, J.~W. and {Alterman}, B.~L. and {Whittlesey}, P.~L. and {Malaspina}, David M. and {Bale}, S.~D. and {Kasper}, J.~C. and {Case}, A.~W. and {Goetz}, K. and {Harvey}, P.~R. and {Korreck}, K.~E. and {MacDowall}, R.~J. and {Pulupa}, M. and {Stevens}, M.~L. and {de Wit}, T. Dudok},
        title = "{Parker Solar Probe Observations of Proton Beams Simultaneous with Ion-scale Waves}",
      journal = {\apjs},
         year = 2020,
        month = may,
       volume = {248},
       number = {1},
          eid = {5},
        pages = {5},
          doi = {10.3847/1538-4365/ab86af},
archivePrefix = {arXiv},
       eprint = {2004.03009},
 primaryClass = {physics.space-ph},
       adsurl = {https://ui.adsabs.harvard.edu/abs/2020ApJS..248....5V}
}

@ARTICLE{alterman18,
       author = {{Alterman}, B.~L. and {Kasper}, Justin C. and {Stevens}, Michael L. and {Koval}, Andriy},
        title = "{A Comparison of Alpha Particle and Proton Beam Differential Flows in Collisionally Young Solar Wind}",
      journal = {\apj},
         year = 2018,
        month = sep,
       volume = {864},
       number = {2},
          eid = {112},
        pages = {112},
          doi = {10.3847/1538-4357/aad23f},
archivePrefix = {arXiv},
       eprint = {1809.01693},
 primaryClass = {astro-ph.SR},
       adsurl = {https://ui.adsabs.harvard.edu/abs/2018ApJ...864..112A}
}

@ARTICLE{hellinger11,
       author = {{Hellinger}, Petr and {Tr{\'a}vn{\'\i}{\v{c}}ek}, Pavel M.},
        title = "{Proton core-beam system in the expanding solar wind: Hybrid simulations}",
      journal = {J.~Geophys.~Res.~(Space Phys.)},
         year = 2011,
        month = nov,
       volume = {116},
       number = {A11},
          eid = {A11101},
        pages = {A11101},
          doi = {10.1029/2011JA016940},
       adsurl = {https://ui.adsabs.harvard.edu/abs/2011JGRA..11611101H}
}

@ARTICLE{marsch87,
       author = {{Marsch}, E. and {Livi}, S.},
        title = "{Observational evidence for marginal stability of solar wind ion beams}",
      journal = {\jgr},
         year = 1987,
        month = jul,
       volume = {92},
       number = {A7},
        pages = {7263-7268},
          doi = {10.1029/JA092iA07p07263},
       adsurl = {https://ui.adsabs.harvard.edu/abs/1987JGR....92.7263M}
}

@ARTICLE{montgomery76,
       author = {{Montgomery}, M.~D. and {Gary}, S.~P. and {Feldman}, W.~C. and {Forslund}, D.~W.},
        title = "{Electromagnetic instabilities driven by unequal proton beams in the solar wind}",
      journal = {\jgr},
         year = 1976,
        month = jun,
       volume = {81},
       number = {16},
        pages = {2743},
          doi = {10.1029/JA081i016p02743},
       adsurl = {https://ui.adsabs.harvard.edu/abs/1976JGR....81.2743M}
}

@ARTICLE{gary85,
       author = {{Gary}, S.~P.},
        title = "{Electromagnetic ion beam instabilities - Hot beams at interplanetary shocks}",
      journal = {\apj},
         year = 1985,
        month = jan,
       volume = {288},
        pages = {342-352},
          doi = {10.1086/162797},
       adsurl = {https://ui.adsabs.harvard.edu/abs/1985ApJ...288..342G}
}

@ARTICLE{gary84,
       author = {{Gary}, S.~P. and {Foosland}, D.~W. and {Smith}, C.~W. and {Lee}, M.~A. and {Goldstein}, M.~L.},
        title = "{Electromagnetic ion beam instabilities}",
      journal = {Phys.~Fluids},
         year = 1984,
        month = jul,
       volume = {27},
       number = {7},
        pages = {1852-1862},
          doi = {10.1063/1.864797},
       adsurl = {https://ui.adsabs.harvard.edu/abs/1984PhFl...27.1852G}
}

@ARTICLE{daughton99,
       author = {{Daughton}, William and {Gary}, S. Peter and {Winske}, Dan},
        title = "{Electromagnetic proton/proton instabilities in the solar wind: Simulations}",
      journal = {\jgr},
         year = 1999,
        month = mar,
       volume = {104},
       number = {A3},
        pages = {4657-4668},
          doi = {10.1029/1998JA900105},
       adsurl = {https://ui.adsabs.harvard.edu/abs/1999JGR...104.4657D}
}

@ARTICLE{matteini13,
       author = {{Matteini}, Lorenzo and {Hellinger}, Petr and {Goldstein}, Bruce E. and {Landi}, Simone and {Velli}, Marco and {Neugebauer}, Marcia},
        title = "{Signatures of kinetic instabilities in the solar wind}",
      journal = {J.~Geophys.~Res.~(Space Phys.)},
         year = 2013,
        month = jun,
       volume = {118},
       number = {6},
        pages = {2771-2782},
          doi = {10.1002/jgra.50320},
       adsurl = {https://ui.adsabs.harvard.edu/abs/2013JGRA..118.2771M}
}

@ARTICLE{yao20,
       author = {{Yao}, Jiansheng and {Gao}, Xinliang and {Chen}, Huayue and {Ke}, Yangguang and {Li}, Yi},
        title = "{The effects of beam proportion on electromagnetic proton/proton instability and associated ion heating: 2D hybrid simulation}",
      journal = {Phys.~Plasmas},
         year = 2020,
        month = feb,
       volume = {27},
       number = {2},
          eid = {022901},
        pages = {022901},
          doi = {10.1063/1.5128744},
       adsurl = {https://ui.adsabs.harvard.edu/abs/2020PhPl...27b2901Y}
}

@ARTICLE{verscharen13,
       author = {{Verscharen}, Daniel and {Chandran}, Benjamin D.~G.},
        title = "{The Dispersion Relations and Instability Thresholds of Oblique Plasma Modes in the Presence of an Ion Beam}",
      journal = {\apj},
         year = 2013,
        month = feb,
       volume = {764},
       number = {1},
          eid = {88},
        pages = {88},
          doi = {10.1088/0004-637X/764/1/88},
archivePrefix = {arXiv},
       eprint = {1212.5192},
 primaryClass = {physics.space-ph},
       adsurl = {https://ui.adsabs.harvard.edu/abs/2013ApJ...764...88V}
}

@ARTICLE{eastwood24,
       author = {{Eastwood}, J.~P. and {Brown}, P. and {Magnes}, W. and {Carr}, C.~M. and {Agu}, M. and {Baughen}, R. and {Berghofer}, G. and {Hodgkins}, J. and {Jernej}, I. and {M{\"o}stl}, C. and {Oddy}, T. and {Strickland}, A. and {Vitkova}, A.},
        title = "{The Vigil Magnetometer for Operational Space Weather Services From the Sun-Earth L5 Point}",
      journal = {Space Weather},
         year = 2024,
        month = jun,
       volume = {22},
       number = {6},
          eid = {e2024SW003867},
        pages = {e2024SW003867},
          doi = {10.1029/2024SW003867},
       adsurl = {https://ui.adsabs.harvard.edu/abs/2024SpWea..2203867E}
}

@ARTICLE{hapgood17,
       author = {{Hapgood}, Mike},
        title = "{L1L5Together: Report of Workshop on Future Missions to Monitor Space Weather on the Sun and in the Solar Wind Using Both the L1 and L5 Lagrange Points as Valuable Viewpoints}",
      journal = {Space Weather},
         year = 2017,
        month = may,
       volume = {15},
       number = {5},
        pages = {654-657},
          doi = {10.1002/2017SW001652},
       adsurl = {https://ui.adsabs.harvard.edu/abs/2017SpWea..15..654H}
}

@INCOLLECTION{hapgood19,
       author = {{Hapgood}, Mike},
        title = "{The Impact of Space Weather on Human Missions to Mars: The Need for Good Engineering and Good Forecasts}",
    booktitle = {The Human Factor in a Mission to Mars},
         year = 2019,
         publisher = {Springer, Cham},
       editor = {{Szocik}, Konrad},
        pages = {69},
          doi = {10.1007/978-3-030-02059-0_4},
       adsurl = {https://ui.adsabs.harvard.edu/abs/2019hfmm.book...69H}
}

@ARTICLE{schwenn06,
       author = {{Schwenn}, Rainer},
        title = "{Space Weather: The Solar Perspective}",
      journal = {Living Reviews in Solar Physics},
         year = 2006,
        month = dec,
       volume = {3},
       number = {1},
          eid = {2},
        pages = {2},
          doi = {10.12942/lrsp-2006-2},
       adsurl = {https://ui.adsabs.harvard.edu/abs/2006LRSP....3....2S}
}

@ARTICLE{miteva23,
       author = {{Miteva}, Rositsa and {Samwel}, Susan W. and {Tkatchova}, Stela},
        title = "{Space Weather Effects on Satellites}",
      journal = {Astronomy},
         year = 2023,
        month = aug,
       volume = {2},
       number = {3},
        pages = {165-179},
          doi = {10.3390/astronomy2030012},
       adsurl = {https://ui.adsabs.harvard.edu/abs/2023Astro...2..165M}
}

@ARTICLE{temmer21,
       author = {{Temmer}, Manuela},
        title = "{Space weather: the solar perspective: An update to Schwenn (2006)}",
      journal = {Living Reviews in Solar Physics},
         year = 2021,
        month = dec,
       volume = {18},
       number = {1},
          eid = {4},
        pages = {4},
          doi = {10.1007/s41116-021-00030-3},
archivePrefix = {arXiv},
       eprint = {2104.04261},
 primaryClass = {astro-ph.SR},
       adsurl = {https://ui.adsabs.harvard.edu/abs/2021LRSP...18....4T}
}

@ARTICLE{rodriguez20,
       author = {{Rodriguez}, L. and {Scolini}, C. and {Mierla}, M. and {Zhukov}, A.~N. and {West}, M.~J.},
        title = "{Space Weather Monitor at the L5 Point: A Case Study of a CME Observed with STEREO B}",
      journal = {Space Weather},
         year = 2020,
        month = oct,
       volume = {18},
       number = {10},
          eid = {e02533},
        pages = {e02533},
          doi = {10.1029/2020SW002533},
       adsurl = {https://ui.adsabs.harvard.edu/abs/2020SpWea..1802533R}
}

@ARTICLE{schrijver15,
       author = {{Schrijver}, Carolus J. and {Kauristie}, Kirsti and {Aylward}, Alan D. and {Denardini}, Clezio M. and {Gibson}, Sarah E. and {Glover}, Alexi and {Gopalswamy}, Nat and {Grande}, Manuel and {Hapgood}, Mike and {Heynderickx}, Daniel and {Jakowski}, Norbert and {Kalegaev}, Vladimir V. and {Lapenta}, Giovanni and {Linker}, Jon A. and {Liu}, Siqing and {Mandrini}, Cristina H. and {Mann}, Ian R. and {Nagatsuma}, Tsutomu and {Nandy}, Dibyendu and {Obara}, Takahiro and {Paul O'Brien}, T. and {Onsager}, Terrance and {Opgenoorth}, Hermann J. and {Terkildsen}, Michael and {Valladares}, Cesar E. and {Vilmer}, Nicole},
        title = "{Understanding space weather to shield society: A global road map for 2015-2025 commissioned by COSPAR and ILWS}",
      journal = {Advances in Space Research},
         year = 2015,
        month = jun,
       volume = {55},
       number = {12},
        pages = {2745-2807},
          doi = {10.1016/j.asr.2015.03.023},
archivePrefix = {arXiv},
       eprint = {1503.06135},
 primaryClass = {physics.space-ph},
       adsurl = {https://ui.adsabs.harvard.edu/abs/2015AdSpR..55.2745S}
}

@ARTICLE{eastwood18,
       author = {{Eastwood}, J.~P. and {Hapgood}, M.~A. and {Biffis}, E. and {Benedetti}, D. and {Bisi}, M.~M. and {Green}, L. and {Bentley}, R.~D. and {Burnett}, C.},
        title = "{Quantifying the Economic Value of Space Weather Forecasting for Power Grids: An Exploratory Study}",
      journal = {Space Weather},
         year = 2018,
        month = dec,
       volume = {16},
       number = {12},
        pages = {2052-2067},
          doi = {10.1029/2018SW002003},
       adsurl = {https://ui.adsabs.harvard.edu/abs/2018SpWea..16.2052E}
}

@ARTICLE{oughton17,
       author = {{Oughton}, Edward J. and {Skelton}, Andrew and {Horne}, Richard B. and {Thomson}, Alan W.~P. and {Gaunt}, Charles T.},
        title = "{Quantifying the daily economic impact of extreme space weather due to failure in electricity transmission infrastructure}",
      journal = {Space Weather},
         year = 2017,
        month = jan,
       volume = {15},
       number = {1},
        pages = {65-83},
          doi = {10.1002/2016SW001491},
       adsurl = {https://ui.adsabs.harvard.edu/abs/2017SpWea..15...65O}
}

\end{document}